\documentclass[12pt]{article}
\usepackage{scicite}
\usepackage{times}

\usepackage[utf8]{inputenc}
\usepackage{amsmath}
\usepackage{graphicx}
\usepackage{hyperref}
\usepackage{xcolor}
\usepackage{gensymb}
\usepackage{lipsum}

\newcommand{\kappaex}{\kappa_{\mathrm{ex}}}
\newcommand{\avdw}{\alpha_{\mathrm{vdW}}}

\newcounter{lastnote}

\title{Nonlinear wave dynamics on a chip}

\author
{Matthew T. Reeves,$^{1\ast}$ Walter W. Wasserman,$^{1\ast}$ Raymond A. Harrison,$^{1}$ \and
Igor Marinkovi\'c,$^{1}$
Nicole Luu,$^{1}$ Andreas Sawadsky,$^{1}$ Yasmine L.
\and Sfendla,$^{1}$ Glen I. Harris,$^{1}$ Warwick P. Bowen,$^{1, 2\dagger}$ Christopher G. Baker,$^{1}$ \vspace{2pt} \and
\normalsize{$^{1}$ARC Centre of Excellence for Engineered Quantum Systems,}\and
\normalsize{School of Mathematics and Physics, University of Queensland, St Lucia, QLD 4072, Australia.}\and
\normalsize{$^{2}$ARC Centre of Excellence in Quantum Biotechnology,}\and
\normalsize{School of Mathematics and Physics, University of Queensland, St Lucia, QLD 4072, Australia.}\vspace{2pt}\and
\normalsize{$^\ast$These authors contributed equally.}\and
\normalsize{$^\dagger$Corresponding author. E-mail:  w.bowen@uq.edu.au.}}

\date{}

\begin{document}

\baselineskip 24pt
\maketitle
\newenvironment{sciabstract}{\begin{quote}\bfseries}{\end{quote}}

\begin{sciabstract}
Shallow water waves are a striking example of nonlinear hydrodynamics, giving rise to phenomena such as tsunamis and undular waves. These dynamics are typically studied in hundreds-of-meter-long wave flumes. Here, we demonstrate a chip-scale, quantum-enabled wave flume. The wave flume exploits nanometer-thick superfluid helium films and optomechanical interactions to achieve nonlinearities surpassing those of extreme terrestrial flows. Measurements reveal wave steepening, shock fronts, and soliton fission -- nonlinear behaviors long predicted in superfluid helium but never previously directly observed. Our approach enables lithography-defined wave flume geometries, optomechanical control of hydrodynamic properties, and orders of magnitude faster measurements than terrestrial flumes. Together, this opens a new frontier in hydrodynamics, combining quantum fluids and nanophotonics to explore complex wave dynamics at microscale.
\end{sciabstract}

\section*{Introduction}

Hydrodynamics governs rich nonlinear behaviors across an extraordinary range of systems and scales -- from Bose–Einstein condensates~\cite{dutton_observation_2001} and cellular transport~\cite{huber2018hydrodynamics} to geophysical flows~\cite{garven1995continental} and interstellar media~\cite{vburkert2006turbulent}. Its realization in miniaturized, chip-scale systems such as polariton condensates~\cite{amo_polariton_2011}, electron fluids~\cite{aharon-steinberg_direct_2022}, superfluid vortex flows~\cite{sachkou_coherent_2019} and microscale mixers~\cite{stroock2002chaotic}, is opening new opportunities to probe these behaviors with unprecedented precision.

\begin{figure*}[t!]
\centering
    \includegraphics[width=\textwidth]{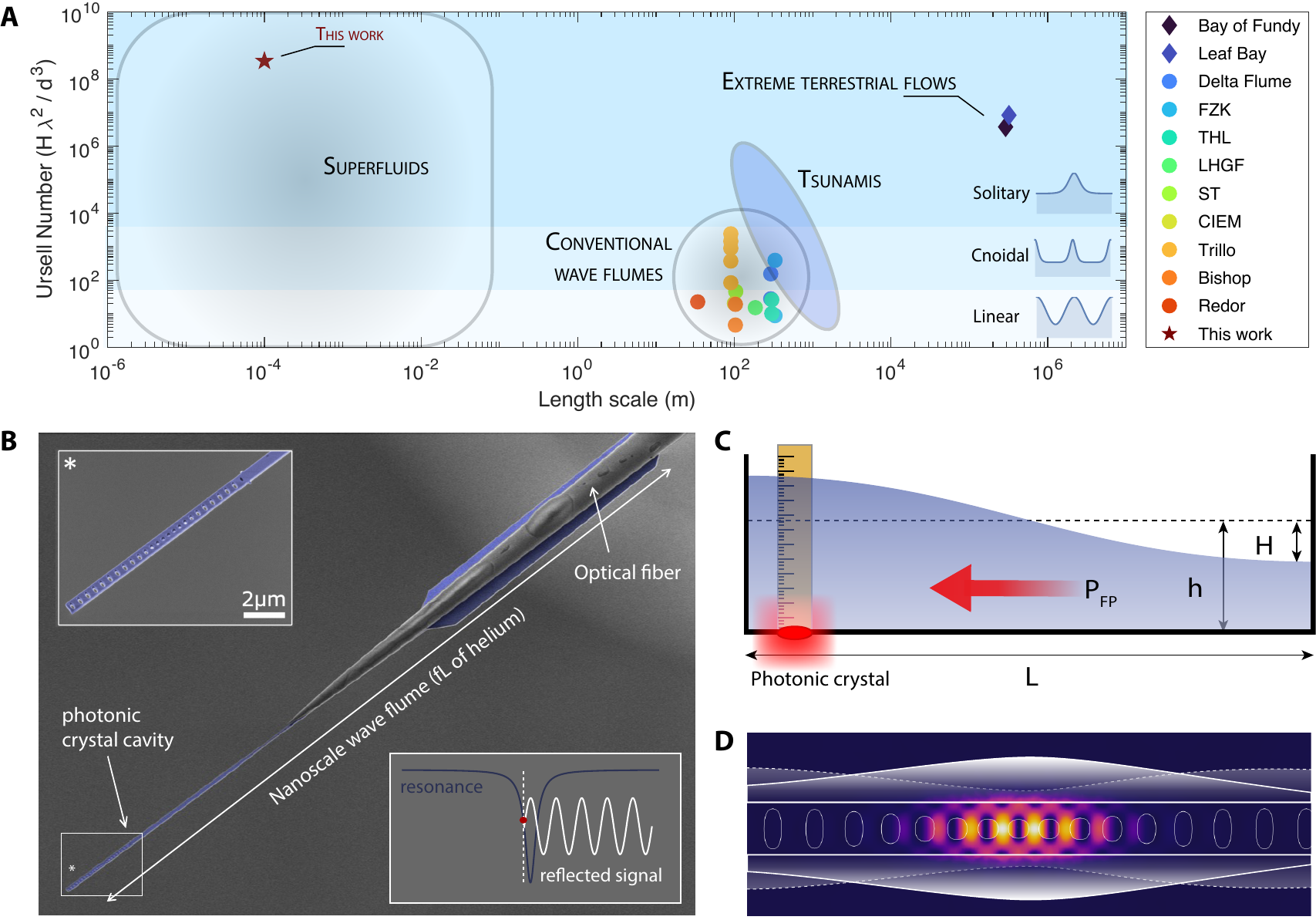}
    \caption{\textbf{Superfluid wave flume}. (\textbf{A}) Landscape of terrestrial shallow wave phenomena and flumes. Conventional wave tanks and flumes cluster in a small region of size and addressable Ursell number, far from extreme terrestrial flows~\cite{schimmels_tsunami_2016}. Superfluid-based approaches open up a significantly broader parameter space~(see Supplementary Materials, Section~\ref{appendixaddressableparameterspace}). (\textbf{B}) False color scanning electron microscope image, showing the silicon wave flume (blue) addressed by an optical fiber taper (gray). 
    Width and thickness (220 nm) being well below its length $L$, the device functions effectively as a one-dimensional channel (\textit{i.e.} a wave flume). {\it Top left inset:} Photonic crystal cavity.  (\textbf{C}) Illustration of the wave flume, including its fundamental sloshing mode. (\textbf{D})  Finite element simulation of the cavity optical mode at $\sim 1598$ nm wavelength. The evanescent component extending outside the cavity senses superfluid wave motion, shown here with exaggerated amplitude (white line).}
    \label{fig:Fig1}
\end{figure*}

Among hydrodynamics, nonlinear shallow water waves are of particular importance. They underpin phenomena ranging from tsunamis and tidal bores to solitons, dispersive shock waves, and integrable turbulence~\cite{hereman2022shallow}. While these phenomena are typically studied in large wave flumes spanning hundreds of meters~\cite{zhang_introduction_2015,trillo_experimental_2016, schimmels_tsunami_2016}, even the largest facilities cannot access the extreme nonlinearities found in nature (Fig.~\ref{fig:Fig1}A). The nonlinearity increases rapidly with decreasing depth~\cite{ursell_long-wave_1953}. As such, chip-scale platforms that use thin fluid films promise far higher nonlinearities -- even with their much shorter length. However, viscosity immobilizes thin films of regular fluid. The exception is superfluid helium, which flows without dissipation even in films just nanometers deep~\cite{scholtz_third_1974}.

Nonlinear 
waves analogous to those in shallow water were predicted in superfluid helium over fifty years ago~\cite{atkins_chapter_1970,ichiyanagi1975nonlinear,huberman1978superfluid}, but their small amplitude and spatiotemporal extent has so far prevented unambiguous observation~\cite{ellis_third_1992}. Here, we overcome this, developing a microscale superfluid helium wave flume that employs a nanophotonic cavity to both drive and probe fluid motion. The wave flume reaches hydrodynamic nonlinearities five orders of magnitude higher than those of the largest terrestrial wave flumes, with values even exceeding those of typical of tsunamis or the world’s most extreme tides~\cite{garrett_tidal_1972}. This allows us to observe a full spectrum of nonlinear wave dynamics, from steepening to near-discontinuous shock fronts and soliton fission into trains of up to twelve solitons. It also allows us to verify the predicted features of hydrodynamics in superfluid helium, including backward steepening~\cite{atkins_chapter_1970} and ‘hot’ solitons that propagate as depressions rather than peaks~\cite{nakajima_solitons_1980, nakajima_solitary_1980}.

The ability to access new regimes of hydrodynamics in an engineerable chip-scale wave flume, together with the prospect of exploiting optomechanical interactions~\cite{bowen2015quantum,aspelmeyer_cavity_2014}, dispersion engineering~\cite{archer_experimental_2020} and rapid measurements over millisecond timescales, opens a path towards resolving long-standing questions in fluid physics and advancing our understanding of complex wave phenomena.

\section*{Superfluid wave flume}

Replicating the capabilities of a large-scale wave flume on a superfluid chip requires a long confining channel 
 to guide the wave during its evolution, a height gauge to monitor the wave amplitude over time, as well as the equivalent of a wave maker to generate large amplitude waves. Fig.~\ref{fig:Fig1}B shows our device architecture, which fulfills these requirements.
It consists of a micro-fabricated silicon beam (blue) with length $L \sim 100~\mu$m, which functions both as an optical and superfluid waveguide. 
A photonic crystal cavity is fabricated at one extremity (see upper inset, Fig.~\ref{fig:Fig1}B). The entire device can be detached and glued to the tip of a tapered optical fiber~\cite{wasserman_cryogenic_2022}, enabling efficient optical coupling even at cryogenic temperatures
(see Supplementary Materials, Section~\ref{methodssectionfabrication}).

When cooled below the superfluid transition temperature in
a $^4$He environment, the device is coated with a self-assembling superfluid helium film whose thickness can be varied from a few monolayers to tens of nanometers~\cite{harris_laser_2016,sachkou_coherent_2019,he_strong_2020,sawadsky_engineered_2023,korsch_phononic_2024}. Here, we select a film thickness of $h=6.7$~nm. 
The film sustains superfluid third sound waves~\cite{harris_laser_2016} that reflect with high efficiency off either end of the silicon beam (see Supplementary Materials Section~\ref{methodsectionacousticeigenmodes}). This forms 
a quasi-one-dimensional wave flume that contains only a few femtoliters of superfluid helium. The fundamental sloshing mode of the flume has wavelength $\lambda= 2L$ (Fig.~\ref{fig:Fig1}C). We find experimentally that its resonance frequency is $\Omega/2\pi=26$ kHz, 
 in good agreement with predictions obtained from finite element simulations ($\Omega_{\mathrm{sim}}/2\pi=27$ kHz, see Supplementary Materials, Section \ref{methodsectionacousticeigenmodes}).

We use optomechanical coupling to readout the superfluid wave dynamics
(Fig.~\ref{fig:Fig1}D). This shifts the cavity resonance frequency in proportion to the wave height, imprinting it onto the output optical intensity \cite{harris_laser_2016, baker_theoretical_2016,sachkou_coherent_2019,he_strong_2020,sawadsky_engineered_2023, korsch_phononic_2024}
(Fig.~\ref{fig:Fig1}B, lower right inset).
To initialize large-amplitude third sound waves,
we use the superfluid fountain pressure $P_{\mathrm{FP}}$, namely superfluids propensity to flow \emph{towards} heat sources~\cite{sawadsky_engineered_2023}.
The cavity thus serves a dual purpose, both as a localized depth gauge and a wave maker (Fig.~\ref{fig:Fig1}C).

The nonlinearity of a shallow water wave is quantified by its Ursell number $\mathrm{Ur} = H \lambda^2 /h^3$, where $H$ is its height. The nanometer-scale thickness of superfluid films and ability to reach large amplitudes provide access to 
Ursell numbers that exceed $10^8$. This is several orders of magnitude larger than is achieved in the world's longest wave flumes~\cite{trillo_experimental_2016}, and comparable to the nonlinearity of the largest terrestrial flows,
all within a sub-millimeter-sized device in a laboratory setting.   
As illustrated in Fig.~\ref{fig:Fig1}A, our superfluid wave flume should therefore be an ideal platform to explore new domains of nonlinear hydrodynamics.

\begin{figure*}[t!]
\centering
    \includegraphics[width=\textwidth]{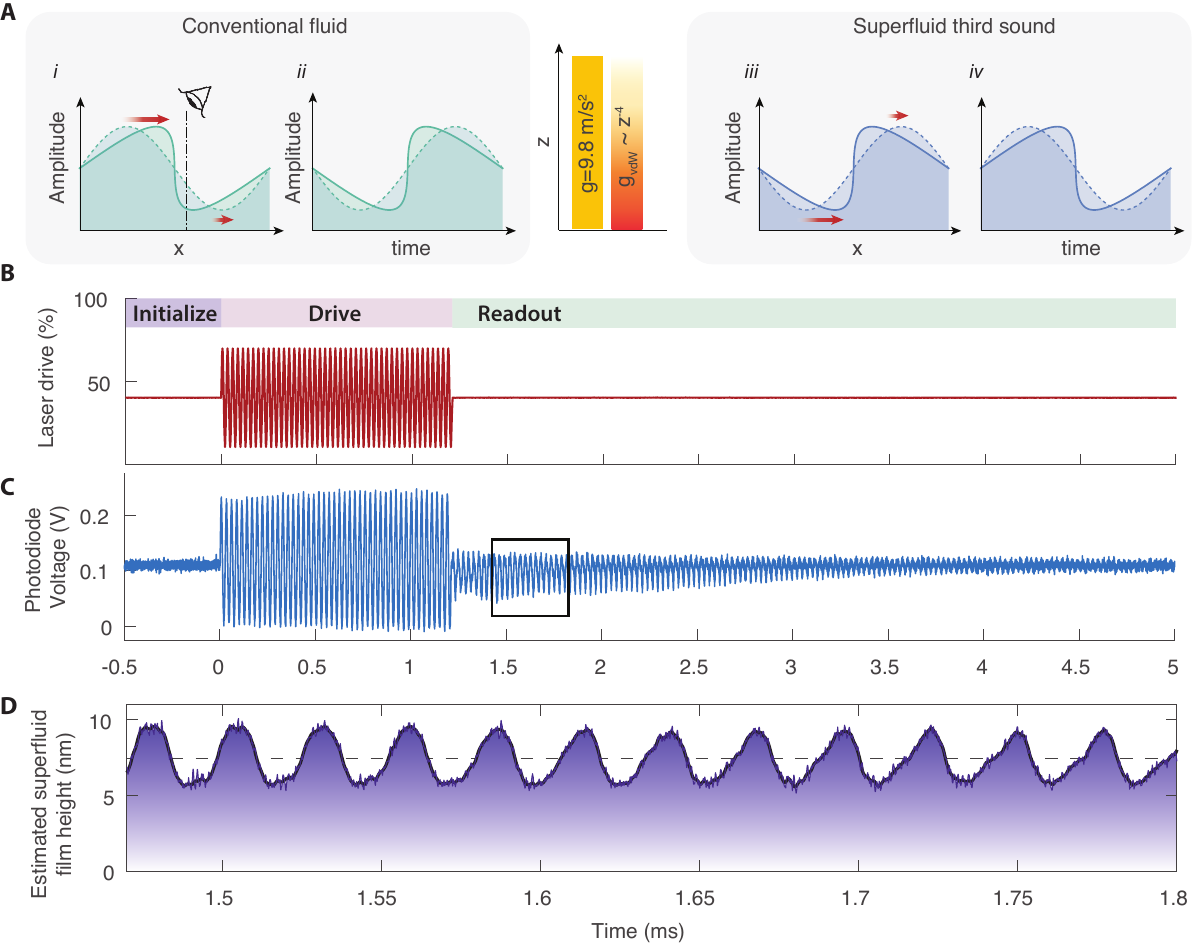}
    \caption{\textbf{Backwards-leaning waves.} (\textbf{A}) \textit{i}: In conventional hydrodynamics wave crests move faster than troughs (red arrows), leading to forward-leaning waves. \textit{ii}: forward-leaning waves as observed by a stationary gauge.  \textit{iii}\&\textit{iv}: reverse behavior is predicted for superfluid third sound waves due to their different effective gravity. 
    (\textbf{B})  Laser drive used to initialize and readout large-amplitude third sound waves. Mean  laser power is kept constant to avoid unwanted thermal drift (see Supplementary Materials, Section~\ref{methodsectionexperimentalsetup}).
   (\textbf{C}) Photodetector voltage measured during wave initialization and readout. 
    (\textbf{D}) Calibrated measurement of wave amplitude versus time, for the section of the ringdown highlighted in (C).}
    \label{fig:Fig2}
\end{figure*}

\section{Third sound waves lean backwards}

A key signature of wave nonlinearity is its  amplitude dependent velocity. In the long wavelength (shallow water) limit  applicable here ($\lambda\gg h$), a wave's velocity is $c\sim \sqrt{g h}$. In conventional gravity waves, since the velocity increases with height, peaks overtake troughs, leading to the well-known phenomenon of (forward) wavefront steepening, and ultimately wave crashing \cite{nayfeh_nonlinear_2004, onorato_rogue_2016}, as illustrated in Fig.~\ref{fig:Fig2}A~\textit{i}\&\textit{ii}.

For superfluid third sound waves the expected behaviour is markedly different~\cite{atkins_chapter_1970}. While the relation $c\sim \sqrt{g h}$ still holds, the gravitational constant $g$ is replaced by the strongly height-dependent van der Waals acceleration $g_{\mathrm{vdW}}=3\alpha_{\mathrm{vdW}}/h^4$~\cite{atkins_chapter_1970}, such that $c\sim \sqrt{3\alpha_{\mathrm{vdW}}/h^3}$. Since $c$ now decreases with wave height, troughs progress faster than crests.
Initially sinusoidal disturbances are then predicted to evolve into backward leaning waves, and ultimately backward `wave breaking'~\cite{atkins_chapter_1970}.  %
This is illustrated in Fig.~\ref{fig:Fig2}A~\textit{iii}. Figure~\ref{fig:Fig2}A~\textit{iv} shows the equivalent (reversed) wave steepening that a stationary observer, such as our static height gauge, would observe over time.

As an initial experiment, we explore this prediction of backward leaning waves.  Figure~\ref{fig:Fig2}B illustrates our procedure to 
produce and read out large amplitude waves. The 
laser is  tuned to the red (i.e. higher wavelength) flank of the  optical resonance. This suppresses optomechanical amplification of wave motion~\cite{sawadsky_engineered_2023, aspelmeyer_cavity_2014}, and  transduces the motion into amplitude modulations  which are detected on a photodetector (see Fig.~\ref{fig:Fig1}B). 
In the first `initialization' phase of an experimental run the laser power is held constant, allowing 
waves from previous experiments to die down. A `drive' phase follows, 
during which the laser is sinusoidally intensity modulated with 80\% peak-to-peak amplitude at the fundamental sloshing mode  frequency of the waveflume  (Fig.~\ref{fig:Fig2}B). 
Optical absorption of the intensity modulated light causes temperature oscillations $\Delta T$  that drive wave motion through the time-varying fountain pressure  $\Delta P=\rho S \Delta T$, with $\rho$ and $S$ respectively the superfluid's density and specific entropy~\cite{sawadsky_engineered_2023}. 
In the final `readout' phase, the laser reverts to its initial constant amplitude to allow a minimally perturbative ringdown measurement of the evolution of the driven third sound waves.

Driving the waves with an incident laser power of $P=240$ nW, we obtain the photodetector output shown in Fig.~\ref{fig:Fig2}C.
In the initialization phase ($-0.5< t< 0$ ms), the surface of the fluid is quiescent, as evidenced by the constant optical transmission. During the drive phase, the photodetector records mainly the applied laser intensity modulation, superimposed with any weaker intensity modulation caused by the wave motion. 
 Finally, the readout phase ($t>1.2$ ms) shows a gradual ringdown of driven wave motion. 

The acquired photodetector trace contains the wave amplitude, but is modified by the Lorentzian response of the optical resonance. We correct this to obtain calibrated measurements of wave amplitude  using the known optomechanical properties of the flume 
(see Supplementary Materials \ref{methodsfilmthicknessestimation}).
Figure~\ref{fig:Fig2}D shows the successful initialization of large-amplitude waves for a section of ringdown shortly after wave initialization. 
Their height  $H \sim 3$ nm is comparable to the mean fluid depth ($h=6.7$ nm), and corresponds to an Ursell number of $4\times10^8$ (solid star, Fig.~\ref{fig:Fig1}A).
From an initially sinusoidal wave profile, we observe the predicted evolution to backward leaning waves during propagation, occurring rapidly over only around twenty wave periods. The magnitude of wave leaning can be precisely characterized through the wave asymmetry parameter~\cite{onorato_rogue_2016}, which evolves from a value near 0 (symmetric) immediately after initialization to -0.6 at  $t=1.8$~ms (see Supplementary Materials). 

\begin{figure*}
\centering
    \includegraphics[width=\textwidth]{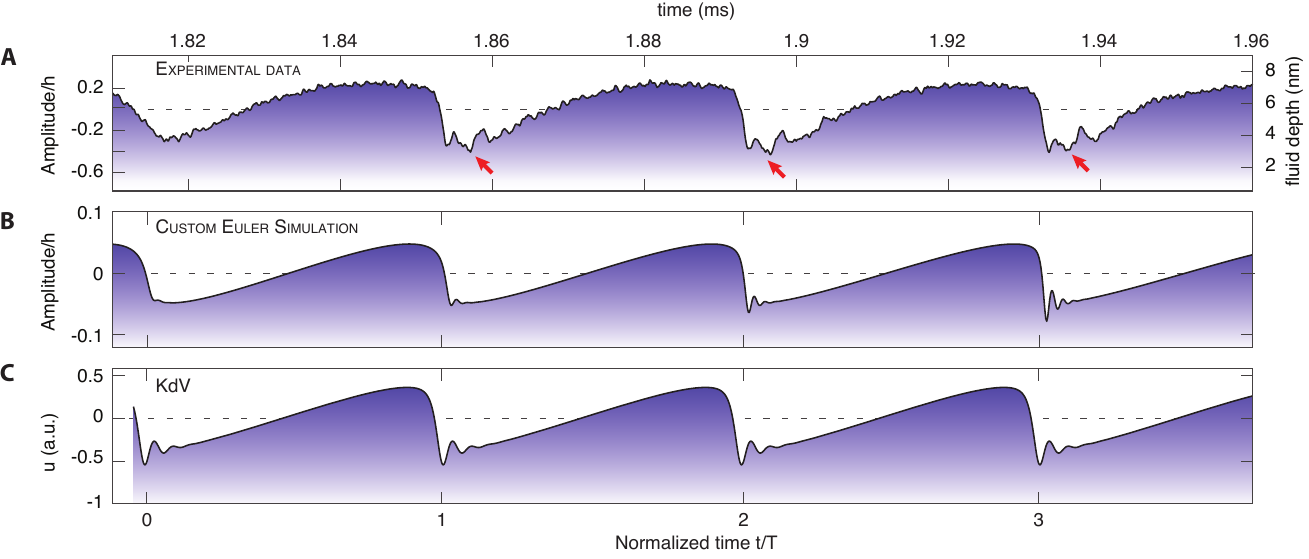}
    \caption{\textbf{Superfluid dispersive shock fronts.} 
    (\textbf{A}) Observation of dispersive shock fronts. Red arrows: signatures of soliton fission. (\textbf{B}) Custom Euler model, including counter-propagation and van der Waals nonlinearities, obtained by evolving a sinusoidal wave initialized with amplitude $H=0.1 h$ (with computation taking $\sim 4$ days on a desktop PC). More details: Supplementary Materials, Section~\ref{methodssectionhydrodynamicmodelling}. (\textbf{C}) KdV solution obtained using Eq.~\eqref{Eq:KdV}.  
   }
    \label{fig:Fig3}
\end{figure*}

\section*{Superfluid films manifest solitons}

The expectation of strong nonlinearity in third sound waves led, over fifty years ago, to the prediction that they could support solitons~\cite{ichiyanagi1975nonlinear,huberman1978superfluid}.
One formation mechanism involves progressive wave leaning, where increasing steepness produces a near-discontinuous shock front followed by soliton fission~\cite{zabusky_interaction_1965}. 
While Fig.~\ref{fig:Fig2}D shows backwards leaning, it does not show the full process
(see Supplementary Materials, Fig.~\ref{fig:ringdownfulllength} for entire ringdown).

To search for shock waves and soliton formation we apply a stronger drive, increasing
%
the incident laser power
fourfold to $P=0.96$ $\mu$W.
Fig.~\ref{fig:Fig3}A plots three oscillation periods, experimentally acquired immediately after the end of the drive (see Supplementary Materials, Fig.~\ref{fig:wavebreakingfulllength} for entire ringdown). 
Compared to the leaning waves in Fig.~\ref{fig:Fig2}D, Fig.~\ref{fig:Fig3}A shows the abrupt drop in amplitude of a shock front. 
The wave steepening time $\tau=\frac{\lambda h}{3 c_3 H}$~\cite{ellis_third_1992} is on the order of the  oscillation period, such that the wave steepens within a single oscillation (between $t=1.8$ and $t=1.85$ ms).

To further understand the observed wave dynamics, we compare them to hydrodynamic simulations. The simulations are computationally challenging due to
the extreme aspect ratio ($L/h \sim 10^{4} \gg 1$) and large wave amplitudes ($H/h \sim 1$). To implement them, we develop a custom hydrodynamic Euler model solver 
which includes both advective and van der Waals nonlinearities, reflection at the flume boundaries, and bidirectional wave propagation (Supplementary Materials, Section~\ref{methodssectionhydrodynamicmodelling}).
 Leveraging the incompressible and irrotational nature of the superflow, the solver calculates the wave's velocity potential $\phi$ 
 and thus provides the entire surface elevation in the wave flume as a function of time, starting from an initially sinusoidal perturbation.  To capture the micrometer-wide spatial extent of the optical mode used for readout, we average the surface elevation across an equivalent region in our simulations. The wave flume geometry increases dispersion by a factor of 35 compared to hydrodynamics alone (see later). We account for this by decreasing the model aspect ratio $h/L$ by the same factor, finding good qualitative agreement with experiment (Fig.~\ref{fig:Fig3}B).

 It is interesting to ask whether a simpler model based on the unidirectional KdV equation might suffice to understand our experimental results, even though this neglects the bidirectional nature of the flow and the higher-order van der Waals nonlinearities. 
The equation  takes the form:
\begin{equation}
    u_t+ \alpha u\,u_x+ \beta u_{xxx}=0,
    \label{Eq:KdV}
\end{equation}
where $u$ is the wave amplitude and the coefficients $\alpha$ and $\beta$ respectively quantify the magnitude of nonlinearity and dispersion~\cite{zabusky_interaction_1965}.  
Solving it, we obtain the dynamics in Fig.~\ref{fig:Fig3}C,
again showing qualitative agreement with experiment.

 The form of Eq.~\eqref{Eq:KdV} allows a qualitative understanding of the physics of shock wave formation.
 As described by Zabusky and Kruskal~\cite{zabusky_interaction_1965}, initially the first two terms dominate. This drives an
overtaking phenomenon: for negative $\alpha$, troughs overtake crests leading to backwards wave steepening. After sufficient steepening, dispersive effects (third term) become important and prevent the appearance of a discontinuous shock. 
Instead, as the shock front evolves, small oscillations appear, growing in amplitude with a shape close to the single solitary wave solution of Eq.~\ref{Eq:KdV}---a phenomenon termed \emph{soliton fission}~\cite{zabusky_interaction_1965,trillo_experimental_2016,redor_experimental_2019}. These solitons propagate with speed linearly proportional to their amplitude, spreading out into a soliton train~\cite{zabusky_interaction_1965, trillo_experimental_2016}.

Fig.~\ref{fig:Fig3}A-C show signatures suggestive of the onset of soliton fission (red arrows, Fig.~\ref{fig:Fig3}A). 
However, these do not evolve into full-blown fission (see full time trace, Supplementary Material Fig.~\ref{fig:wavebreakingfulllength}).
Fission is suppressed because high-frequency oscillations decay faster than the fundamental sloshing mode, a long-observed phenomenon~\cite{bergman_third_1971,atkins_chapter_1970,generazio_attenuation_1984}   that our analytical models confirm (see Supplementary Materials~\ref{sectionatkinsdispersionrelationanddamping} and \ref{sectionsuppevaporativedamping}).  
%

\begin{figure*}[t]
\centering
    \includegraphics[width=\textwidth]{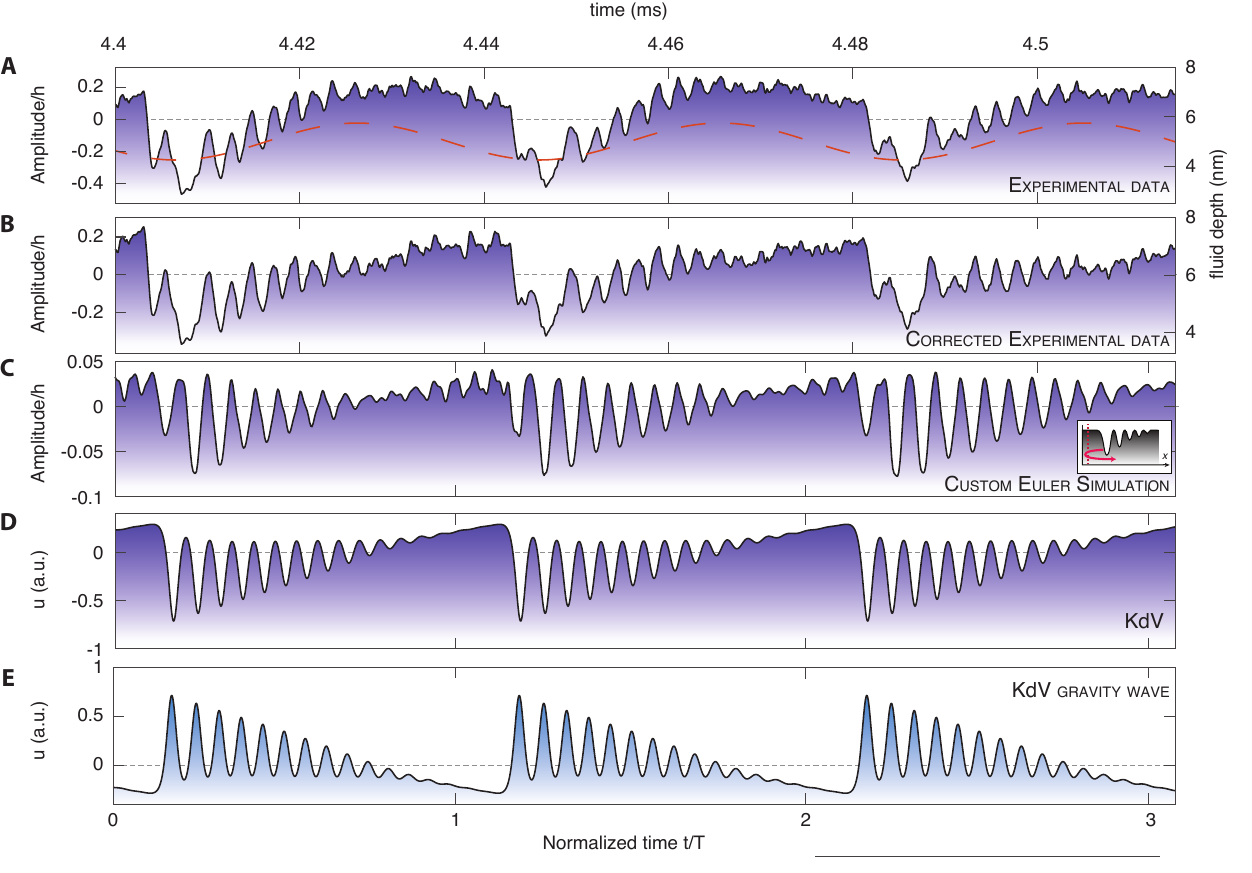}
    \caption{\textbf{Multi-soliton fission.} (\textbf{A}) Observation of multisoliton fission, superimposed over a sinusoidal modulation at the drive frequency with amplitude $H=1.5$ nm (dashed red line). (\textbf{B})~Corrected experimental data, with sinusoidal background subtracted. (\textbf{C}) Custom Euler model.  
    (\textbf{D}) Third sound KdV solution obtained using Eq.~\eqref{Eq:KdV}. (\textbf{E}) Gravity wave KdV solution obtained using Eq.~\eqref{Eq:KdV}, with reversed nonlinear coefficient $\alpha$.}
    \label{fig:Fig4}
\end{figure*}

To observe soliton fission, we double the injected laser power to $P=1.9$ $\mu$W and extend the drive period to 3~ms. Figure~\ref{fig:Fig4}A shows an experimental  trace of the wave dynamics recorded immediately after the drive. The  dynamics include a superimposed sinusoidal modulation due to the optical amplitude modulation near the end of the drive period (dashed red sinusoid). To compare the experiments with theory we subtract this in Fig.~\ref{fig:Fig4}B. Both figures exhibit a train of twelve solitons. It has only recently become possible to observe soliton trains of this size in hundred meter-scale wave flumes~\cite{trillo_experimental_2016}.
 In striking contrast to these regular shallow-water solitons, which manifest as propagating peaks (Fig.~\ref{fig:Fig4}E), our superfluid solitons exist as propagating depressions below the equilibrium fluid depth. 
This counter-intuitive property has been long-predicted for third-sound solitons~\cite{nakajima_solitary_1980,nakajima_solitons_1980}.
Since the normal fluid component is viscously clamped and does not move, and the temperature is determined by the ratio of normal to superfluid, such solitons have locally increased temperature and are referred to as `hot solitons'~\cite{nakajima_solitary_1980,nakajima_solitons_1980}. 

  Figures~\ref{fig:Fig4}C\&D show the predictions of our custom Euler solver and the KdV solver, respectively, in both cases showing full-blown soliton fission and good agreement with our measurements (see also Supplemental Movie S1). 
Here,  the improved modelling capabilities of the custom method are evident, both in the better prediction of height for the first two solitons, and in the faithful modeling of high-frequency components absent in the KdV solution once the soliton train dies away to low amplitude.

Modelling predicts that the first soliton in the train should have the largest amplitude~\cite{zabusky_interaction_1965, trillo_experimental_2016}, consistent with our KdV equation model. Figures~\ref{fig:Fig4}B\&C appear to contradict this, with the second soliton dip having the largest amplitude. This
is  an artifact due to the slight offset of our optical height gauge from the boundary of the wave flume. As illustrated in the inset of Fig.~\ref{fig:Fig4}C, each soliton passes the gauge twice, first as an incoming wave and then again upon reflection. After reflection, outgoing solitons interfere with subsequent incoming solitons in the train, modulating their perceived amplitude
(Fig.~\ref{fig:Fig4}C and Supplemental Movie S1).

The number of solitons emerging from the fission process should scale with the ratio of nonlinear to dispersive length scales~\cite{trillo_experimental_2016}.
With only hydrodynamic wave dispersion,
in excess of 200 solitons are expected
(see Supplementary Materials, Section~\ref{sectionappendixdispersion}).  The smaller number we observe indicates that much larger dispersion is at play. We therefore perform a detailed dispersion analysis, including normal hydrodynamic dispersion, anomalous dispersion due to surface tension, and coupling to superfluid temperature oscillations (see Supplementary Materials Section~\ref{sectionappendixdispersion}). 
We find that the \emph{acoustic} Bragg grating formed by the photonic crystal holes of our optical cavity introduces normal dispersion~\cite{sfendla_extreme_2021,korsch_phononic_2024}.
This dominates  the contribution of surface tension. The ultimate effect is to increase the normal dispersion by a factor of $\sim$35, so that
the observed physics is well reproduced in purely hydrodynamic simulations with an aspect ratio $h/L$ of only 300:1.
The engineered acoustic dispersion serves to stabilize  micrometer-long `hot' solitons, rather than the 100~nm-scale `cold' solitons predicted 
when surface tension dispersion dominates~\cite{nakajima_solitary_1980}. Without engineered dispersion, `hot' solitons are only predicted for few monolayer films and with nanometer-scale length~\cite{nakajima_solitons_1980} (see Supplementary Materials, Section ~\ref{sectionappendixinfluenceofdispersionandnonlinearity}).


The dispersive effects of Bragg mirrors near the edge of the band-gap are well known. They are widely used in photonics for dispersion compensation, slow light generation and sensing~\cite{skolianos_slow_2016}, and have recently been employed in a large-scale wave flume for water wave amplification and filtering \cite{archer_experimental_2020}.  In our platform, the unique combination of anomalous capillary dispersion and engineerable normal dispersion 
provides the capacity to fully tailor dispersion all the way from normal to anomalous regimes.


\section*{Discussion}

The observation of nonlinear wave dynamics in superfluid films has been an outstanding goal for many decades~\cite{atkins_chapter_1970,ichiyanagi1975nonlinear,huberman1978superfluid}. Although soliton-like signatures have previously been inferred using bolometric techniques~\cite{kono_anomalous_1981,mckenna_observation_1990}, limited spatiotemporal resolution precluded unambiguous detection. Subsequent efforts using capacitive probes failed~\cite{ellis_third_1992},  likely hindered by low resolution and sensitivity as well as the inability to drive large-amplitude waves near the breaking threshold. 

Our use of optomechanical interactions in a nanophotonic cavity overcomes these challenges, providing high spatiotemporal resolution,  sub-picometer sensitivity, and strong driving.
High-speed readout is achieved via the radiation pressure interaction~\cite{aspelmeyer_cavity_2014}, while fountain-pressure actuation is independently optimised~\cite{sawadsky_engineered_2023}. This resolves a key incompatibility in typical optomechanical systems, where strong radiation-pressure forcing requires high-quality-factor cavities, but fast measurement demands low quality factors~\cite{sawadsky_engineered_2023}.  Moreover, optomechanical dynamics, even in liquids~\cite{kaminski_ripplon_2016 , kashkanova_superfluid_2017, he_strong_2020, harris_laser_2016}, are generally well described by one or a few harmonic modes, with nonlinearities treated perturbatively. The wave steepening, shock formation, and soliton fission we observe lie well outside this regime. As such, our work represents a qualitative shift in optomechanics experiments from perturbative corrections to fundamentally nonlinear dynamics.

Our wave flume offers unique opportunities to explore nonlinear hydrodynamics. Optomechanical interactions afford new capabilities for dynamic control, including tunable gain~\cite{sawadsky_engineered_2023} and dissipation~\cite{harris_laser_2016}, and even real-time control of dispersion~\cite{zhang_optomechanical_2021}. 
The capacity of superfluid helium to self-assemble over nanofabricated substrates~\cite{harris_laser_2016} allows arbitrary wave flume geometries and dispersion profiles to be defined lithographically. Moreover, unlike ordinary fluids, the effective gravity of superfluid helium films 
can be widely tuned by varying film thickness. This enables a continuous transition between gravity- and capillary-wave regimes, and therefore between fundamentally different energy and wave-action cascades~\cite{falcon_observation_2007}.

The effective gravity is orders of magnitude stronger than the Earth’s gravitational acceleration, resulting in wave speeds comparable to those in open water despite nanometer-scale depths. This dramatic compression of spatial and temporal scales accelerates experimental cycles by a factor of a million compared to terrestrial wave flumes~\cite{trillo_experimental_2016}, opening new avenues for data-driven discovery -- from machine learning of governing equations to predictive modeling of turbulent and multiscale flows~\cite{kochkov_machine_2021,zhu_review_2022,mojgani2024interpretable}.
Combined with the ability to reach far higher nonlinearities than other hydrodynamic experiments, these capabilities open new paths to address long-standing questions in nonlinear wave physics, including soliton interactions~\cite{el_kinetic_2005,el_soliton_2021},  modulational instabilities~\cite{benjamin_disintegration_1967}, energy and wave-action cascades~\cite{deike_energy_2014,abdurakhimov_bidirectional_2015}, and the emergence of structures such as breathers in near-integrable systems~\cite{kivshar_dynamics_1989}.

\section*{Acknowledgements}
{\bf Funding:} This work was primarily funded by the U.S. Army Research Office through grants number W911NF17-1-0310 and W911NF-23-1-0117. It also received support through  
 the Australian Research Council Centre of Excellence for Engineered Quantum Systems (EQUS, Project No. CE170100009), is in part 
 based upon work supported by the Defense Advanced Research Projects Agency (DARPA) under Agreement No. HR00112490528.
C.G.B,  G.I.H, and M.T.R respectively acknowledge Australian Research Council Fellowships DE190100318 \& FT240100405, DE210100848, and DE220101548. This work was performed in part at the Queensland node of the Australian National Fabrication Facility, a company established under the National Collaborative Research Infrastructure Strategy to provide nano and micro-fabrication facilities for Australia’s researchers. The authors acknowledge the facilities, and the scientific and technical assistance, of the Microscopy Australia Facility at the Centre for Microscopy and Microanalysis (CMM), the University of Queensland. The authors acknowledge insightful discussions with Andrea Al\`{u} and Seunghwi Kim.
{\bf Author contributions} M.T.R., R.A.H., N.L., W.W.W., Y.L.S. and C.G.B. performed the numerical simulations of waves and devices.  W.W.W. and C.G.B. acquired the experimental data. I.M. and A.S. fabricated the devices, with assistance from G.I.H.. R.A.H. and G.I.H. contributed to the experimental setup. C.G.B., W.P.B. and M.T.R. wrote the manuscript. C.G.B. and W.P.B. conceived and supervised the project. All authors discussed and analyzed the results and provided feedback on the manuscript.
{\bf Competing interests:}
The authors declare no competing interests.
{\bf Data and materials availability:}
All data needed to evaluate the conclusions in the paper are present in the paper and the Supplementary Materials. Additional simulation files and scripts will be accessible from the online Zenodo data repository.

\section*{Supplementary Materials}

\begin{flushleft}

Materials and Methods

Figs. S1 to S17

Table S1 

References (54) to (77) 

Movie S1

\end{flushleft}

\bibliographystyle{science}
\bibliography{bibliography}

\clearpage
\newpage



\renewcommand{\thetable}{S\arabic{table}}%
\renewcommand{\figurename}{Fig.}
\renewcommand{\thefigure}{S\arabic{figure}}
\renewcommand{\thepage}{S\arabic{page}}

\setcounter{section}{0}
\setcounter{table}{0}
\setcounter{equation}{0}
\setcounter{page}{1}
\setcounter{figure}{0}

\begin{center}

\section*{Supplementary Materials for Nonlinear wave dynamics on a chip}

Matthew T. Reeves,$^{1\ast}$ Walter W. Wasserman,$^{1\ast}$ Raymond A. Harrison,$^{1}$ \vspace{-10pt}\\
Igor Marinkovi\'c,$^{1}$ Nicole Luu,$^{1}$ Andreas Sawadsky,$^{1}$ Yasmine L.\vspace{-10pt}\\
Sfendla,$^{1}$ Glen I. Harris,$^{1}$ Warwick P. Bowen,$^{1, 2\dagger}$ Christopher G. Baker,$^{1}$ \\
\vspace{5pt}
\normalsize{$^\ast$These authors contributed equally.}\\
\normalsize{$^\dagger$Corresponding author. E-mail:  w.bowen@uq.edu.au.}

\end{center}



\baselineskip 14pt


\section{Comparison with existing experimental landscape}

\begin{figure}[h]
 \centering
     \includegraphics[width=\textwidth]{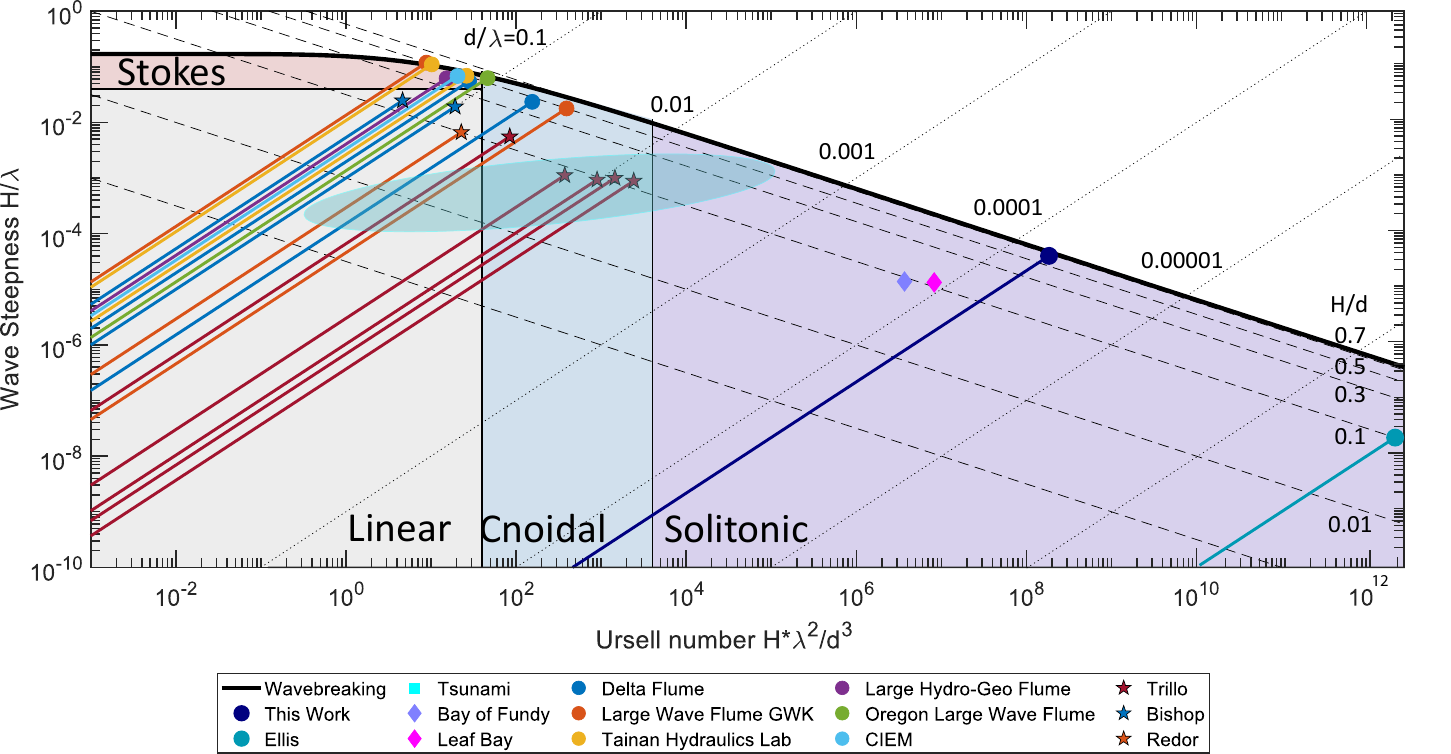}
     \caption{\textbf{Comparison with existing experimental landscape.} This figure presents large scale wave tanks, wave flumes, natural wave phenomena and superfluid helium thin film experiments, plotted as a function of Ursell number $U=\frac{H\,\lambda^2}{d^3}$~\cite{ursell_long-wave_1953} and wave steepness $H/\lambda$.  The thick solid black line materialises the onset of the wave breaking regime~\cite{mehaute_introduction_1976}. Oblique dotted lines mark lines of constant shallowness $d/\lambda$, while oblique dashed lines delineate regions of constant relative wave height $H/d$.  Referenced works are Trillo~\cite{trillo_experimental_2016}; Redor~\cite{redor_experimental_2019}; Bishop~\cite{bishop_measuring_1987} and Ellis~\cite{ellis_quantum_1993, ellis_third_1992,ellis_observation_1989}.}
     \label{Figuresuppstateoftheartsteepnessvsursell}
 \end{figure}

Figure~\ref{Figuresuppstateoftheartsteepnessvsursell} presents a non-exhaustive overview of the experimental landscape, with different large scale wave tank,  wave flume and superfluid thin film experiments, alongside natural wave phenomena, plotted as a function of Ursell number $U=\frac{H\,\lambda^2}{d^3}$~\cite{ursell_long-wave_1953} (with $d$ the fluid depth) and wave steepness $H/\lambda$. This representation is commonly used to highlight the regions of validity of analytical wave theories~\cite{hedges1995regions}. The thick solid black line  materialises the onset of the wave breaking phenomenon, where the particle velocity at the crest becomes larger than the wave velocity~\cite{mehaute_introduction_1976}. The wave breaking criterion takes different forms in deep, intermediate and shallow water regimes; it imposes a maximum relative wave height $H/d<0.78$ for gravity waves in the solitary wave regime~\cite{mehaute_introduction_1976} relevant here. 
As the wave height $H$ appears in both the steepness and the Ursell number, increasing the wave amplitude $H$ corresponds to moving diagonally upward in the log log plot of Fig.~\ref{Figuresuppstateoftheartsteepnessvsursell}.  Since the wave height $H$ can generally be continuously varied from zero to some maximal value $H_{\mathrm{max}}$ in  experiments,   the accessible regimes in different experiments are given by oblique lines in this representation, with the symbol marking the position of the highest quoted wave height $H_{\mathrm{max}}$.

Classical experiments typically cluster in the linear and cnoidal regimes, with $U$ in the $10^0$ to $10^3$ range. Natural phenomena such as tidal resonances and tsunamis are also included for comparison. The broad spread for tsunami waves is representative of their properties in the deep ocean (left-most boundary, where $U_r\sim 1$) and during shoaling (right boundary, where steepness and Ursell number increase significantly)~\cite{madsen_solitary_2008}.
As discussed in the main text, superfluid experiments occupy a unique parameter space, as their viscosity-free flow enables extreme aspect ratios and Ursell numbers to be achieved. 
Pioneering superfluid third sound resonators developed by Ellis et al. (turquoise dot) confined nanometer thick superfluid $^4$He films within  centimeter-scale capacitively readout and actuated third sound resonators~\cite{ellis_high-resolution_1983, ellis_quantum_1993, ellis_third_1992,ellis_observation_1989}, corresponding to extreme aspect ratios $\lambda/d$ in excess of $10^6$. Despite the extreme Ursell number achieved, in excess of $10^{12}$, these experiments suffered from limitations in their temporal and spatial resolution, as well as drive strength which are overcome here.
Indeed, their measurement time was slow compared to the period of the wave, precluding instantaneous measurement of the surface profile, as presented here. Instead a power spectrum of the resonator is acquired over many oscillations, providing access only to averaged wave properties, such as the resonance frequency and linewidth.  The large detector capacitor size~\cite{ellis_high-resolution_1983} also limited the achievable spatial resolution. Secondly, their actuation strength was limited. Indeed, these relied on capacitive actuation,  where the dielectric superfluid is pulled into the high electric field region between the capacitor plates. This effect is weak, due to helium's small permittivity ($\varepsilon_r=1.058$) and large capacitor plate separation~\cite{donnelly_observed_1998, harris_proposal_2020},  leading to a small maximum relative wave height $H/d$,  well below the wave breaking limit.

\subsection{Addressable parameter space}
\label{appendixaddressableparameterspace}

This on-chip platform for nonlinear hydrodynamics promises the opportunity to explore nonlinear wave dynamics in a uniquely large parameter space. In terms of Ursell number, arrays of wave flumes with dimensions ranging from the micrometer scale up  to several millimeters can be fabricated on a single silicon chip. This allows the wavelength $\lambda$ to be scaled by three orders of magnitude. What's more, the superfluid film thickness $d$ is another widely tunable parameter. As we have previously experimentally shown,  the film thickness $d$ can be tuned by one order of magnitude, from $\sim$3 nm to $\sim$30 nm, with sub-nm precision~\cite{harris_laser_2016,sawadsky_engineered_2023,baker_theoretical_2016}, thereby scaling the ratio $d/\lambda$ by one order of magnitude. We can moreover generate and readout waves with  amplitude $H$ ranging from sub-monolayer~\cite{baker_theoretical_2016,sawadsky_engineered_2023} to the wave-breaking limit. Combined, these three degrees of freedom enable our approach to cover an extraordinarily large parameter space---spanning over 9 orders of magnitude in Ursell number (from $10^2$ up to values in excess of $10^{11}$), thereby covering the entire blue and purple regions of the parameter space of Fig.~\ref{Figuresuppstateoftheartsteepnessvsursell} with a single chip.

In this work, we have investigated the free evolution of highly nonlinear waves after their initialization. Building upon these results, in future work  one can leverage capabilities which are unique to our experimental system: the ability to control gain, dissipation and viscosity through interaction with the laser field. As has been shown, the optomechanical interaction can laser-cool~\cite{harris_laser_2016} or amplify~\cite{sawadsky_engineered_2023} surface waves, and can even control their dispersion~\cite{zhang_optomechanical_2021}, while at high laser powers, evaporative effects can introduce an effective viscosity (see Supplementary Material sections~\ref{sectionsuppevaporativedamping} and \ref{sectionatkinsdispersionrelationanddamping}). For thicker films ($h>\sim50$ nm), surface tension becomes the dominant restoring force, thus also providing the ability to widely tune the Eötvös/Bond number all the way from gravity to capillary waves, through the intermediate regime of gravity-capillary waves.  
Our on-chip superfluid wave flume architecture can therefore extend to the study of highly nonlinear driven hydrodynamic flows, integrable turbulence, soliton gases~\cite{redor_experimental_2019}, energy cascades in capillary wave turbulence~\cite{deike_energy_2014,abdurakhimov_bidirectional_2015} and modulational instabilities~\cite{benjamin_disintegration_1967}.

We note moreover that the self-assembling nature of the superfluid films makes this a very flexible platform, allowing experiments customized wave flume or wave tank geometries. Controlled corrugation of the flume's outer boundary may further be used to  control the dispersion (see section~\ref{sectionappendixdispersion}), noting that the periodicity of the photonic and phononic crystals need not be identical as is the case here.
Going further, this approach is compatible with the use of distributed height sensors, as well as the use of plasmonic or small mode-volume dielectric resonances to further improve the spatial resolution.

\section{Experimental details}

\subsection{Device fabrication}
\label{methodssectionfabrication}

\begin{figure}[t!]
    \centering
    \includegraphics[width=\textwidth]{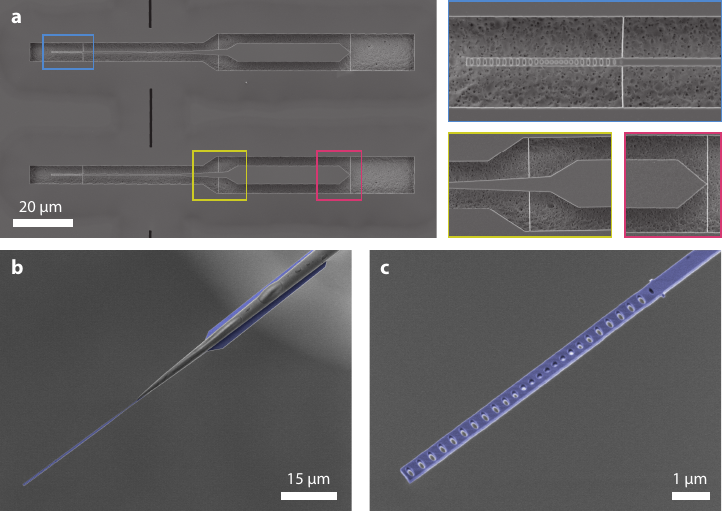}
    \caption{\textbf{Device fabrication. a} Left: Scanning electron micrograph top-view of two fabricated devices on a silicon on insulator chip. The devices have been released with vapor-phase HF (see \ref{methodssectionfabrication}) and are only held in place in three positions (blue, red and yellow boxes) with sub-100 nm wide tethers. Right: zoom-ins of the boxed regions.  \textbf{b}  False color scanning electron microscope image  showing the silicon superfluid wave flume (blue) glued to a silica optical fiber taper (gray). Droplets of NOA 86H UV-curing epoxy are visible on the top side of the silica fiber. \textbf{c} Zoomed-in view showing the silicon photonic crystal cavity at the end of the device. The silicon waveguide is approximately 500 nm wide and 220 nm thick. Small remnants of the broken off tethers are visible in the top right hand side.  }
    \label{fig:supplementaryfabrication}
\end{figure}

The silicon devices are fabricated from a silicon-on-insulator wafer (220~nm silicon device layer on 3~$\mu$m silica (SiO$_2$) buried oxide, on a 675~$\mu$m silicon handle wafer) purchased from Waferpro. Devices are defined on the silicon substrate through electron-beam lithography, using ARP6200.09 positive resist and reactive ion etching. A subsequent hydrofluoric acid (HF) vapour etch selectively removes the silica buried oxide from under the devices, leaving them attached only via small tethers to the rest of the silicon layer, as shown in the electron microscope images in Fig.~\ref{fig:supplementaryfabrication}(a).  

At this point a conically tapered fiber is coated with a small amount of glue (NOA 86H from Norland Optical Adhesives) and aligned to a device in order to achieve optical coupling~\cite{groblacher_highly_2013}.  Once coupling is achieved the glue is partially cured using a UV lamp (Polytec UV-LC 5) 
for 3 minutes. The fiber is then lifted away from the chip, tearing the device off with it, as shown in Fig.~\ref{fig:supplementaryfabrication}(b).  The fiber with the device is then mounted to a 25mm diameter SEM stub using Stycast epoxy (2850FT) and the UV glue is then fully cured by heating to 125\degree C in an oven. As detailed in Ref.~\cite{wasserman_cryogenic_2022}, this procedure produces robust adhesion even at millikelvin temperatures, with the bond remaining intact even after multiple thermal cycles.

\subsection{Experimental setup}
\label{methodsectionexperimentalsetup}

The mounted device (see Fig.~\ref{fig:supplementaryfabrication}(b)) is placed  within a superfluid-tight sample chamber attached to the mixing chamber plate of a Bluefors dilution refrigerator (base temperature 10 mK). Helium-4 gas can be added to the sample chamber via a small capillary in order to vary the thickness of the superfluid film in-situ~\cite{he_strong_2020, sawadsky_engineered_2023}.  The sample chamber contains some alumina nanopowder which increases the effective surface area by $\sim$10 m$^2$, allowing for improved film thickness control and stability~\cite{ellis_observation_1989}. 

Laser light is sent to the device through the conically tapered fiber, and the reflected light from the optical cavity is directly measured with a photodetector. By measuring the reflected power when the laser is slightly detuned from the optical resonance with a power-meter, 
the average coupling efficiency can be determined. For the lower Q mode used in this work, the average (one-way) coupling efficiency is 30\%.  For the higher Q mode present the coupling efficiency is 15\%. Devices are characterised using a widely tunable (1480-1630 nm) Santec TSL-550 laser.  As discussed in the main text, superfluid waves are generated by modulating the laser through its LF (Low-frequency) amplitude modulation port. This port accepts voltages ranging from -2V to 0V, with a corresponding modulation depth of 50\%/V. We apply a voltage to the LF port of the form $V_{\mathrm{LF}}=V_{\mathrm{offset}} + V_{\mathrm{mod}} \sin\left(\Omega t\right)$, with $V_{\mathrm{offset}}$ on the order of -1V. 
The amplitude modulation is therefore symmetric around $V_{\mathrm{offset}}$ and keeps the mean laser intensity constant, as shown in Fig.~\ref{fig:Fig2} of the main text. This prevents any low-frequency changes to the device temperature.  These temperature changes are undesirable here, as they can cause changes to the mean thickness of the film due to the superfluid fountain pressure effect~\cite{sawadsky_engineered_2023}. 

\subsection{Estimating the mean superfluid depth $h$ and the superfluid wave amplitude.}
\label{methodsfilmthicknessestimation}

\begin{figure}
    \centering
    \includegraphics[width=\textwidth]{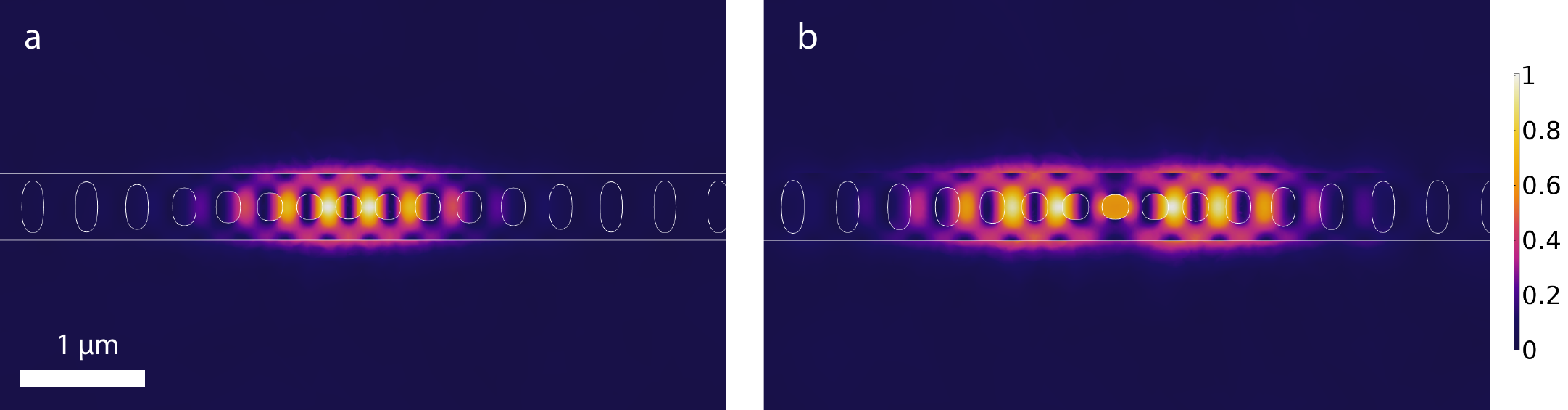}
    \caption{\textbf{Photonic crystal cavity optical resonances.} Finite element simulation, showing a top view of the photonic crystal cavity at the extremity of the superfluid wave flume. Left: optical resonance at 1524 nm; right: optical resonance at 1598 nm. Surface color displays the electric field norm, normalized to $\max(||\vec{E}||)=1$.}
    \label{fig:photoniccrystalopticalmodes}
\end{figure}

To estimate the static superfluid film thickness,  we monitor the shift of the optical mode as helium is added to the sample chamber, as described in detail in Refs~\cite{he_strong_2020,sawadsky_engineered_2023}. This shift is due to the change in effective refractive index caused by the presence superfluid helium, and the observed frequency shift $\Delta \omega_0$ can then be related to a given superfluid film thickness increase $\Delta x$  by using the simplified approximation for the optical frequency shift per unit superfluid film thickness $G = \Delta \omega_0 /\Delta x$  given in Ref.~\cite{baker_theoretical_2016}:

\begin{equation}
    G = \dfrac{\Delta \omega_0}{\Delta x} = \dfrac{-\omega_0}{2} \dfrac{\int_{\mathrm{interface}} (\varepsilon_{\mathrm{sf}}-1) |\vec{E}(\vec{r})|^2 d^2 \vec{r}}{\int_{\mathrm{all\, space}} \varepsilon_{r} |\vec{E}(\vec{r})|^2 d^3 \vec{r}},
\end{equation}
where $\varepsilon_{\mathrm{sf}}=1.058$ is the relative permittivity of superfluid helium~\cite{donnelly_observed_1998}.  As this expression neglects the exponential decay of the evanescent field away from the silicon interface, it is only reasonable for thin-films as considered here.  For substantially thicker films, the decay of the field should be considered. 
The fabricated crystal cavity supports two optical resonances in our laser wavelength range, shown in Fig.~\ref{fig:photoniccrystalopticalmodes}.

These are experimentally measured to be at $\lambda_0=1524$ nm and $\lambda_0=1598$ nm at cryogenic temperatures. For these optical modes, the frequency shift per unit displacement $G/2\pi$ is found from finite element simulations to be 6.2~GHz/nm and  5.6~GHz/nm respectively.  The mean superfluid depth $h$ in our experiments simply corresponds to the measured optical frequency shift during helium injection divided by the optomechanical coupling rate, $h=\Delta \omega_0/G$~\cite{he_strong_2020,sawadsky_engineered_2023}. 

To estimate the wave amplitude $\eta$, we must describe how oscillatory changes in the superfluid film thickness are imprinted on  the intensity of the laser light reflected from the optical cavity. This requires knowledge of the optical mode properties, as well as the optomechanical coupling rate $G$. The optical resonance can be fitted with a Lorentzian mode shape with known baseline $b$ and minimum $b\, (1-a)$, whose reflection profile $R$ is given by~\cite{aspelmeyer_cavity_2014}:
\begin{equation}
    R = b \left(1-\dfrac{a}{1+\left(\dfrac{\bar{\Delta}}{\kappa/2}\right)^2} \right).
    \label{eqn:IntroLorentzian}
\end{equation}
Here $\kappa=\kappa_i+\kappaex$ is the total optical cavity loss rate. It  corresponds to the sum of the intrinsic loss rate $\kappa_i$ (which describes losses such as scattering and absorption) and extrinsic loss rate $\kappaex$ (which describes losses to the coupling waveguide)~\cite{aspelmeyer_cavity_2014}, see Table~\ref{Table_physical_params} for their experimental values.  The parameter $a$ is  the normalized  contrast of the optical mode (where zero corresponds to no dip in the reflection spectrum, and 1 corresponds to full contrast, \textit{i.e.} \emph{critical coupling}~\cite{aspelmeyer_cavity_2014}). It is given by $a=1-\left(\frac{1-\frac{\kappaex}{\kappa_i}}{1+\frac{\kappaex}{\kappa_i}}\right)^2$~\cite{baker2013chip} and its value is fixed for a given optical device and is measured experimentally.  $\bar{\Delta}=\omega_L - \omega_{cav}$ is the effective detuning between the laser frequency $\omega_L$ and the wave-modulated cavity resonance $\omega_{cav}=\omega_0+G \eta$; $\omega_0$ is the equilibrium cavity resonance frequency when the wave amplitude is zero, corresponding to a detuning $\Delta_0=\omega_L-\omega_0$. The effective detuning therefore takes the form $\bar{\Delta}=\Delta_0 - G \eta$.  Combining this relationship with Eq.~\eqref{eqn:IntroLorentzian}, the wave amplitude  $\eta$  can be estimated:
\begin{equation}
    \eta = \dfrac{1}{G} \left(\Delta_0 \pm\dfrac{\kappa}{2} \sqrt{\dfrac{a}{1-\frac{R}{b}} -1} \right).
    \label{eqn:IntroInvertedLorentzian}
\end{equation}
We note that the inversion of the reflection function \ref{eqn:IntroLorentzian} in the form of \ref{eqn:IntroInvertedLorentzian} is not single-valued. 
Indeed, because of the Lorentzian response of the cavity, there are two detunings (on the blue and red side of resonance) that lead to the same optical reflection value, and hence two corresponding wave amplitude solutions. By restricting the measurements to a single side of resonance, and keeping the wave amplitude $\eta$ small enough such that the effective detuning $\bar{\Delta}$ does not change sign (\textit{i.e.} the laser always remains on the same side of resonance) ensures that the relationship between reflected intensity $R$ and wave amplitude $\eta$ remains bijective, that is, that a given optical reflection level corresponds to a unique wave amplitude. The laser should therefore always stay on the same side of resonance, and also not go far out of resonance, as the transduction sensitivity then becomes weak. This criterion imposes a maximal wave amplitude $\eta_{\max}\sim \kappa/G$. For this reason, 
we purposefully employ the optical resonance with the lowest optical quality factor, at $\lambda_0=1598$ nm (Fig.~\ref{fig:photoniccrystalopticalmodes}~(b)), with $Q\simeq 2000$. Its lower optical Q (and hence larger $\kappa$) enables a greater dynamic range and the effective transduction of the large amplitude superfluid waves generated here.

\subsection{Influence of drive asymmetry}
\label{subsectioninfluenceofdriveasymmetry}

\begin{figure}[h]
    \centering
    \includegraphics[width=\textwidth]{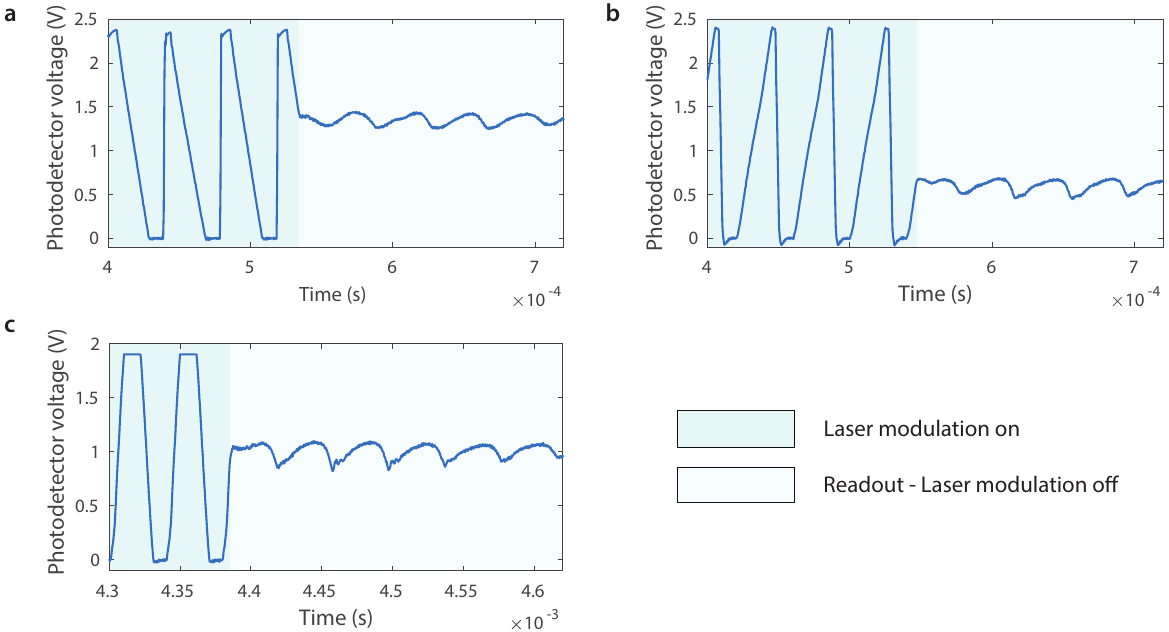}
    \caption{\textbf{Influence of drive asymmetry.} \textbf{a}; \textbf{b} and \textbf{c}: third sound waves respectively generated with \textit{inverse sawtooth}, \textit{sawtooth} and \textit{triangular} waveforms. Measurements taken with the laser blue-detuned from the cavity resonance. }
    \label{fig:driveasymmetry}
\end{figure}

To rule out any influence of the laser actuation on the observed wave asymmetry (and confirm its nonlinear hydrodynamic origin) we measure the wave evolution after initialization through laser drives of varying asymmetry.  Following the approach outlined in the main text, we replace the sinusoidal laser intensity modulation in the drive phase by a `\textit{Ramp}' waveform in our waveform generator.
This waveform allows us to continuously vary the drive asymmetry, through the \textit{`Symmetry'} parameter which is defined as the percentage that the rising period takes up in the whole period. (This parameter can take any value between 0 and 100\%, with 0\% corresponding to a backward leaning \textit{inverse sawtooth} wave, 50\% to a symmetric \textit{triangular wave} and 100\% to a forward leaning \textit{sawtooth} wave).
The results of this are shown in Fig.~\ref{fig:driveasymmetry}. While the magnitude of the generated waves varies slightly, the generated third sound wave asymmetry is independent of the asymmetry of the drive, confirming its hydrodynamic origin.

\section{Full ringdown time traces}

In this section, we show full ringdown time traces acquired in the wave steepening regime (Fig.~\ref{fig:ringdownfulllength}), in the wave breaking regime (Fig.~\ref{fig:wavebreakingfulllength}), and in the soliton fission regime (Fig.~\ref{fig:solitonfissionfulllength}). These respectively correspond to  Figs. ~\ref{fig:Fig2}, \ref{fig:Fig3} and \ref{fig:Fig4} of the main text, in which only short portions of the ringdown were shown for clarity.

\subsubsection{Wave steepening regime} 
\begin{figure}
    \centering
    \includegraphics[width=\textwidth]{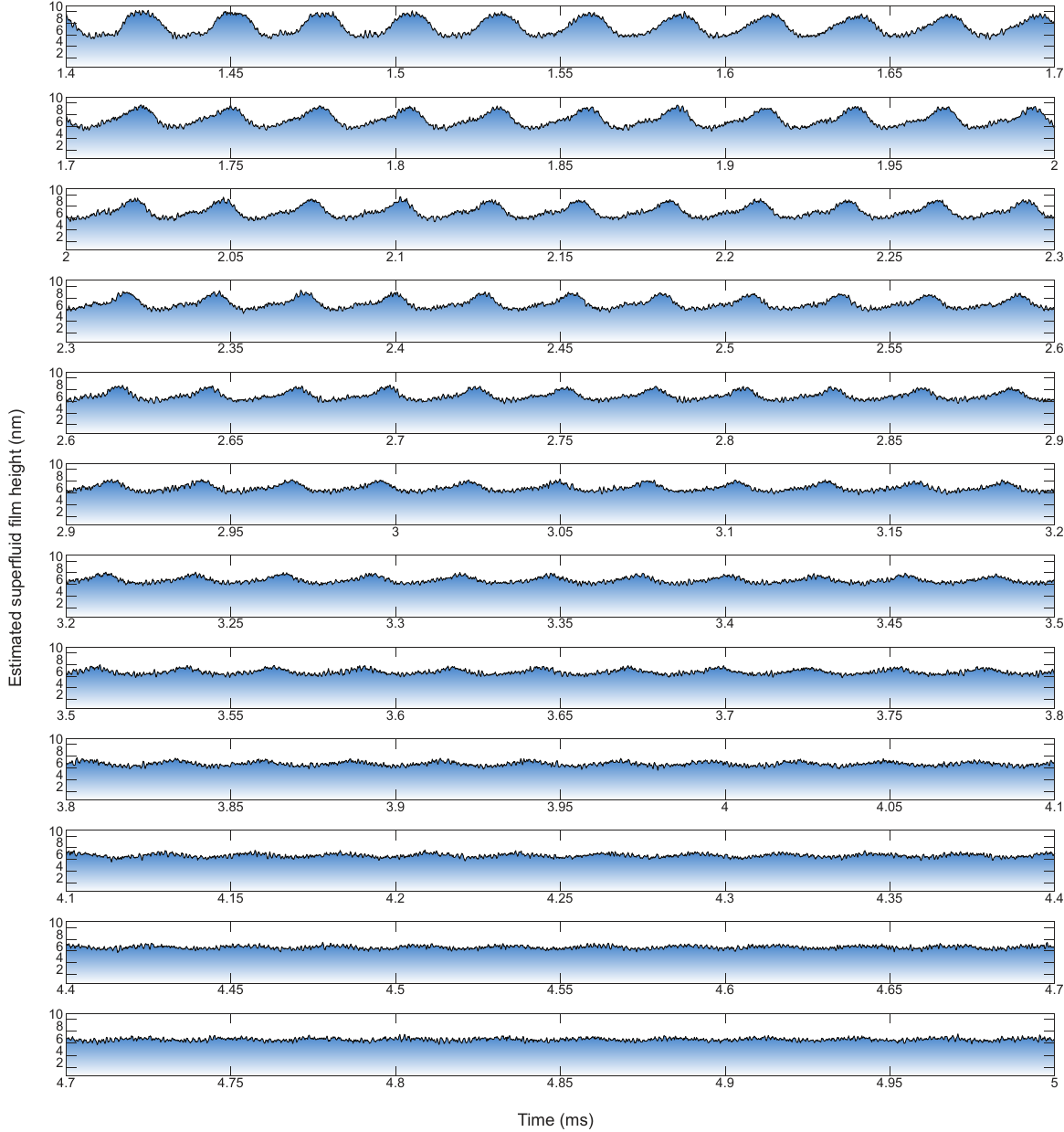}
    \caption{\textbf{Ringdown measurement in the wave steepening regime.} Experimental data showing the estimated superfluid film height during the ringdown measurement shown in Fig.~\ref{fig:Fig2} of the main text.} 
    \label{fig:ringdownfulllength}
\end{figure}

\begin{figure}
    \centering
    \includegraphics[width=.8\textwidth]{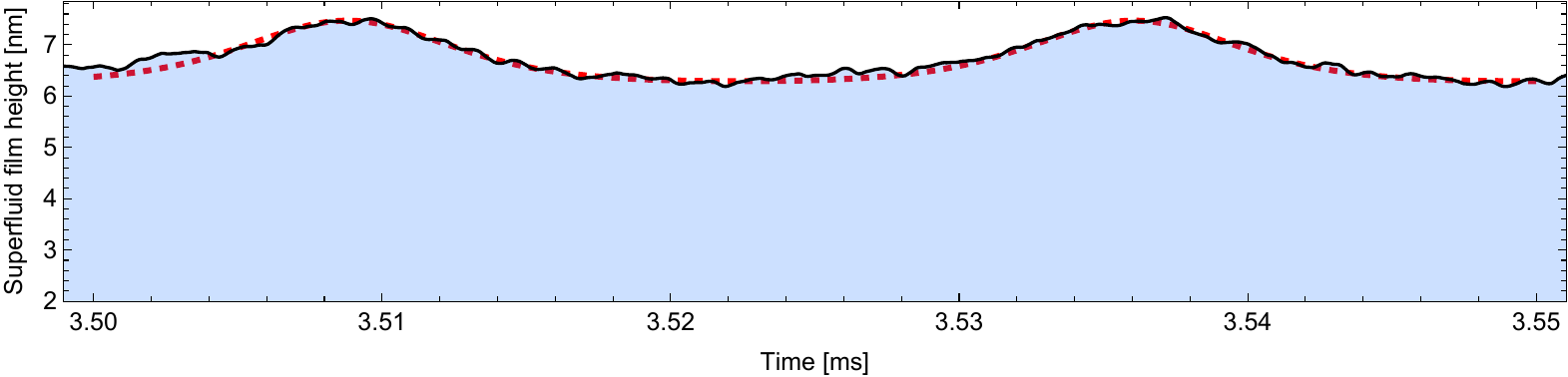}
    \caption{\textbf{Cnoidal wave profile.} Solid black line: experimental data. Dashed red line: cnoidal wave fit. The experimental data is smoothed here using a simple moving average filter of width 25.  Fitting parameters: trough elevation $\eta_2=6.29$ nm; wave height $H=1.18$ nm; elliptic parameter $m=0.949$. }
    \label{fig:cnoidal wave fit}
\end{figure}

Figure~\ref{fig:ringdownfulllength} shows a long ringdown time trace, plotting the wave evolution over $\sim130$ periods after initialization. The initial stages (first five panels), show clear signs of backward wave steepening. As discussed in the main text, higher frequency components of the wave experience more attenuation. This means the high spatial frequency components in the steepened wavefront are preferentially damped, ultimately limiting wave steepening and maintaining more power in the fundamental mode of oscillation.
At intermediate timescales (panels 7, 8, 9), before the wave has decayed into the noise, the wave asymmetry is reduced, and the wave profile can be seen to deviate from the sinusoidal profile of the drive with which it was initialized. In this regime, the wave profile can be well approximated by a cnoidal wave~\cite{onorato_rogue_2016}, characteristic of surface gravity waves in the long wavelength limit.
The cnoidal wave has a surface elevation  $\eta(x,t)$ defined by:
\begin{equation}
    \eta(x,t)=\eta_2+ H cn^2\left(2\, K \left(m\right) \frac{x-c_3\, t}{\lambda} ; m\right),
    \label{Eqn:cnoidalwaveprofile}
\end{equation}
where $\eta_2$ is the trough elevation, $H$ the wave height, $c_3$ the speed of third sound, $\lambda$ the wavelength,  $K(m)$  the complete elliptic integral of the first kind and $cn(u, m)$ the Jacobi elliptic function depending on the elliptic parameter $m$ (where $0<m<1$ and $m=0$ corresponds to a cosine function).
Figure~\ref{fig:cnoidal wave fit} shows a fit of the measured wave profile  between 3.5 and 3.55 ms with Eq.~(\ref{Eqn:cnoidalwaveprofile}). The wave deviates significantly from a sinusoidal profile, and is well approximated by a cnoidal wave with elliptic parameter $m\simeq 0.95$
characterized by sharper peaks and flatter troughs.

\subsubsection{Wave breaking/dispersive shockwave regime}

\begin{figure}
    \centering
    \includegraphics[width=\textwidth]{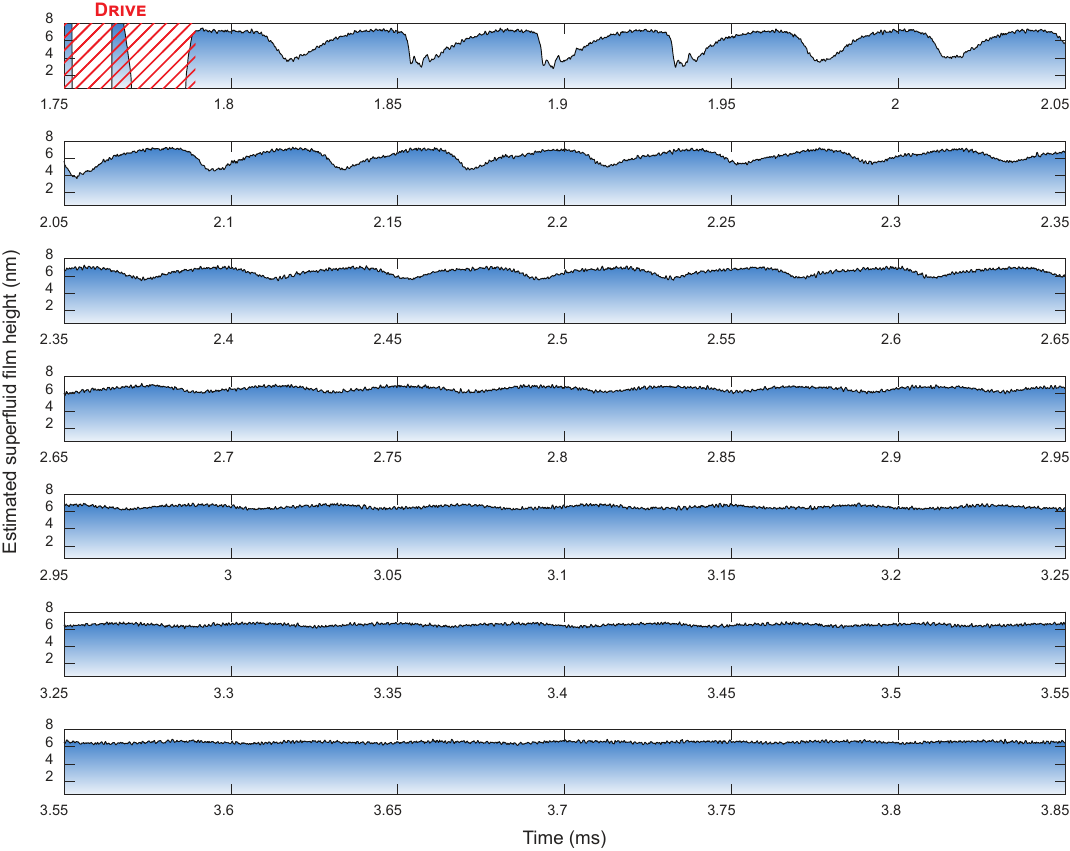}
    \caption{\textbf{Ringdown measurement in the wave breaking regime.} Experimental data showing the estimated superfluid film height during the ringdown measurement shown in Fig.~\ref{fig:Fig3} of the main text. The hatched red region at the beginning of the trace marks the end of the drive phase which initializes the wave. As the laser is amplitude modulated during the drive, the inversion procedure converting photodetector voltage to estimated superfluid film height is not meaningful in this region.}
    \label{fig:wavebreakingfulllength}
\end{figure}

Figure~\ref{fig:wavebreakingfulllength} shows a long ringdown time trace, plotting the wave evolution over $\sim 50$ periods of oscillation after the wave is initialized, in the wave breaking regime shown in Fig.~\ref{fig:Fig3} of the main text. The wave initially steepens in the first period of  free evolution after the drive is switched off ($t\sim 1.8$ to $1.9$ ms). This leads to the formation of a near vertical 
dispersive shock front. The dispersive nature of the front is evidenced by the high frequency ripples formed in the leading edge of the wave trough.  Indeed,  dispersive (or non-dissipative) shock waves are identified by density ripples or oscillatory wave trains whose front propagates faster than the local speed of sound in the medium~\cite{simmons_what_2020, onorato_rogue_2016}, while dissipative shock waves, in contrast,  are characterised by a  steep but smooth change in the density\cite{simmons_what_2020, mossman_dissipative_2018}.
Here the wave steepening time $\tau=\frac{\lambda}{3 c_3}(h/H)$~\cite{ellis_third_1992} is on the order of the period of oscillation, such that the wave steepens within a single oscillation in the wave flume (between $t=1.8$ and $t=1.85$ ms).
As in Fig.~\ref{fig:ringdownfulllength}, steepening and asymmetry are then gradually reduced by frequency-dependent attenuation. Compared to Fig.~\ref{fig:ringdownfulllength}, the wave is more rapidly damped, which we ascribe to laser induced damping, as the incident laser power has been increased fourfold between the two traces. 
As discussed in more detail in section~\ref{sectionsuppevaporativedamping}, we evaluate the resulting evaporative damping of the acoustic wave.
Assuming 1\% of the dissipated power in the device is absorbed as heat, we find that  $\sim 1.7$\% of the mass in the superfluid wave flume is evaporated every oscillation period, providing an upper bound on the quality factor of 59, broadly consistent with the experiments.

\subsubsection{Soliton fission regime}

\begin{figure}
    \centering
    \includegraphics[width=\textwidth]{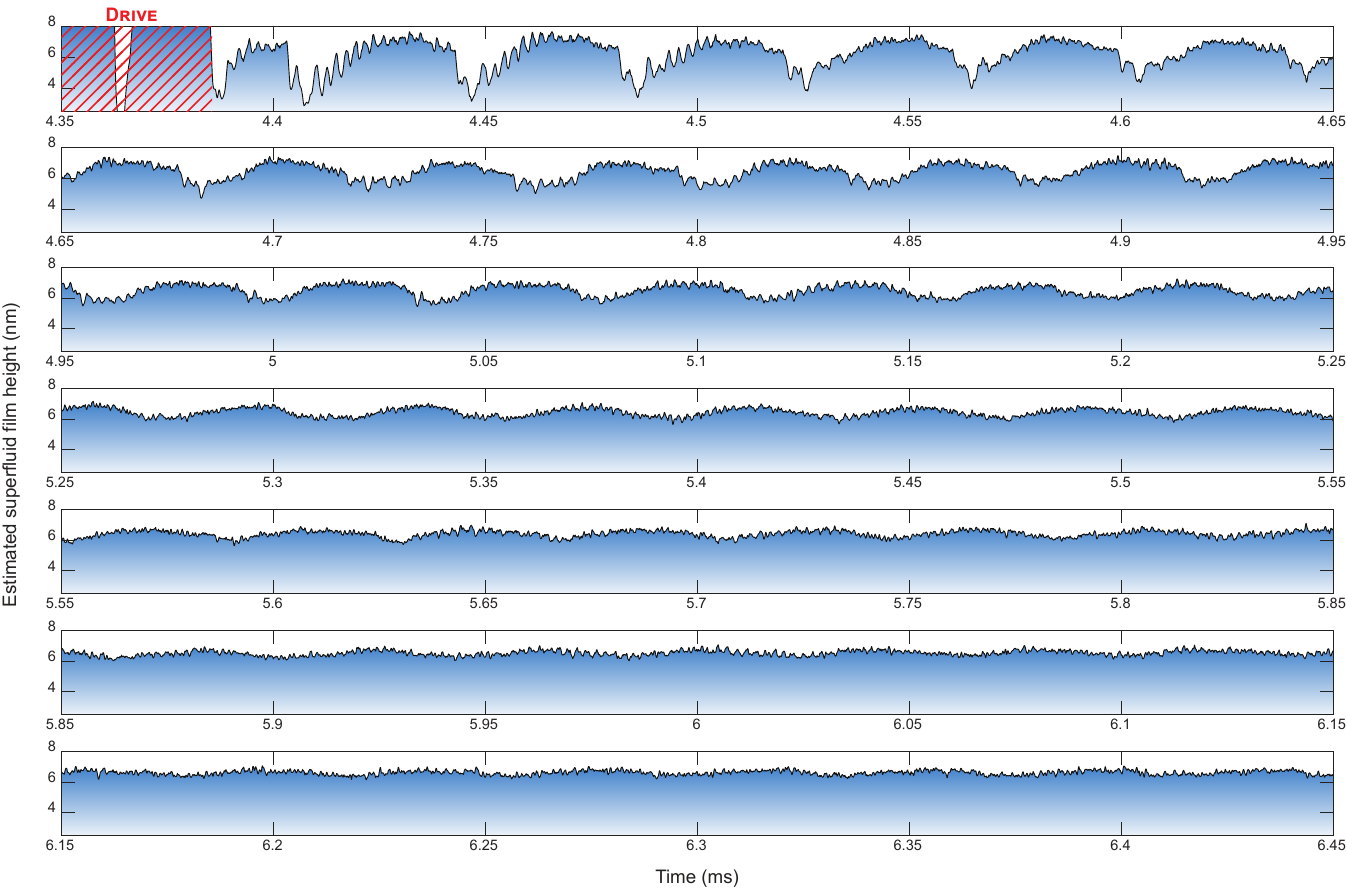}
    \caption{\textbf{Ringdown measurement in the soliton fission regime.}  Experimental data showing the estimated superfluid film height during the ringdown measurement shown in Fig.~\ref{fig:Fig4} of the main text.  The hatched red region at the beginning of the trace marks the end of the drive phase which initializes the wave. As the laser is amplitude modulated during the drive, the inversion procedure converting photodetector voltage to estimated superfluid film height is not meaningful in this region. }
    \label{fig:solitonfissionfulllength}
\end{figure}

Figure~\ref{fig:solitonfissionfulllength} shows a long ringdown time trace, showing the wave evolution over $\sim 50$ periods after initialization, in the soliton fission regime shown in Fig.~\ref{fig:Fig4} of the main text. 
Solitons are generated during the drive phase, and therefore immediately present at the end of the drive.
 At longer timescales, more high spatial frequency noise remains in comparison to the time trace at lower excitation (compare for example panels 5 and 6 in Fig.~\ref{fig:solitonfissionfulllength} and Fig.~\ref{fig:wavebreakingfulllength}), possibly indicative of a regime of soliton gas~\cite{redor_experimental_2019}.

\section{Acoustic eigenmodes}
\label{methodsectionacousticeigenmodes}

In this section, we detail the simulation of the acoustic eigenmodes of the superfluid wave flume. These are calculated here through finite element method simulations in the linear regime. While these linear simulations do not account for nonlinear wave behaviour such as wave breaking and soliton fission, they provide estimates of the superfluid mode eigenfrequencies which are in very good agreement with the experimentally measured values. 
The challenge here is to obtain the eigenmode solutions where the superfluid motion is confined to the 2D outer surface of a non trivial 3D geometry. Indeed, the silicon wave flume is essentially a rectangular prism, but with lateral tapering and photonic crystal holes. To address this, we follow the approach outlined in Ref.~\cite{sawadsky_engineered_2023} and note that the superfluid helium flow in the third sound wave can be considered inviscid, irrotational and incompressible. As such it is a potential flow  (where the velocity potential $\phi$ verifies $\Delta \phi =0$~\cite{mehaute_introduction_1976}), and in the limit of small wave amplitude, the out-of-plane deflection of the superfluid surface $\eta\left(\vec{r},t\right)$ obeys the simple wave equation:
\begin{equation}
    \left(\nabla^2-\frac{1}{c_3^2}\frac{\partial^2}{\partial t^2}\right) \eta=0
\end{equation}
Here $c_3$ is the speed of sound, which neglecting the influence of surface tension, takes the form
$c_3=\sqrt{3 \frac{\rho_s}{\rho}\frac{\avdw}{d^3}}$ \cite{baker_theoretical_2016}. Assuming a separable time-harmonic standing wave solution, of the kind $\eta\left(\vec{r},t\right)=\eta\left(\vec{r}\right) e^{i\Omega t}$, leads to the Helmholtz equation for the spatial mode profile $\eta(\vec{r})$:
\begin{equation}
    \left(\nabla^2+k^2\right)\eta(\vec{r})=0, 
    \label{eqn:helmholtz}
\end{equation}
where $k^2=\frac{\Omega^2}{c_3^2}$ and the displacement profile $\eta(\vec{r})$ is defined on the 2D surface of the 3D photonic crystal resonator geometry. 
As detailed in Ref.~\cite{sawadsky_engineered_2023}, Eq.(\ref{eqn:helmholtz}) can be solved with the help of finite element modelling software (Comsol Multiphysics). 
Figure~\ref{fig:suppFEMacousticmodes} shows the fundamental mode of oscillation of the superfluid film wave flume:  Fig.~\ref{fig:suppFEMacousticmodes}(a) provides an artistic illustration, while Fig.~\ref{fig:suppFEMacousticmodes}(b) shows a finite element method (FEM)  simulation of the fundamental mode of oscillation of the superfluid wave tank, calculated in the linearized regime with the exact device geometry.
This side-by-side comparison illustrates the position of the reflective boundaries (dashed lines). It also illustrates how the photonic crystal cavity serves a dual purpose: as an optomechanical height gauge measuring the superfluid film thickness and as a wave maker driving wave motion through the superfluid fountain effect (black arrow).
As can be seen in the FEM simulation in Fig.~\ref{fig:suppFEMacousticmodes}(b), the displacement amplitude only depends on the $x$ coordinate along the length of the resonator, and is independent of the transverse coordinate $y$, such that the 1D approximation used in the hydrodynamic Euler model in section~\ref{methodssectionhydrodynamicmodelling} is justified. 

\subsubsection{Boundary conditions}
Acoustic boundary conditions at both extremities of the resonator are `free' boundary conditions, where the out-of-plane displacement of the superfluid surface is not constrained, as illustrated in Fig.~\ref{fig:suppFEMacousticmodes}(a).
The `free' boundary condition on the photonic crystal side (left side in Fig.~\ref{fig:suppFEMacousticmodes}(b)) is unambiguous. The precise nature of the boundary condition on the right side is less evident, due to the presence of the glued tapered fiber in the experiments (see Fig.~\ref{fig:supplementaryfabrication})(b)). We verify the `free' boundary condition on the right side through two means. First, the simulated eigenfrequency with \emph{`free-free'} boundary conditions is in good agreement with the experimentally measured frequency of the fundamental sloshing mode of the flume, as discussed in the main text. 

\begin{figure}
    \centering
    \includegraphics[width=.85\textwidth]{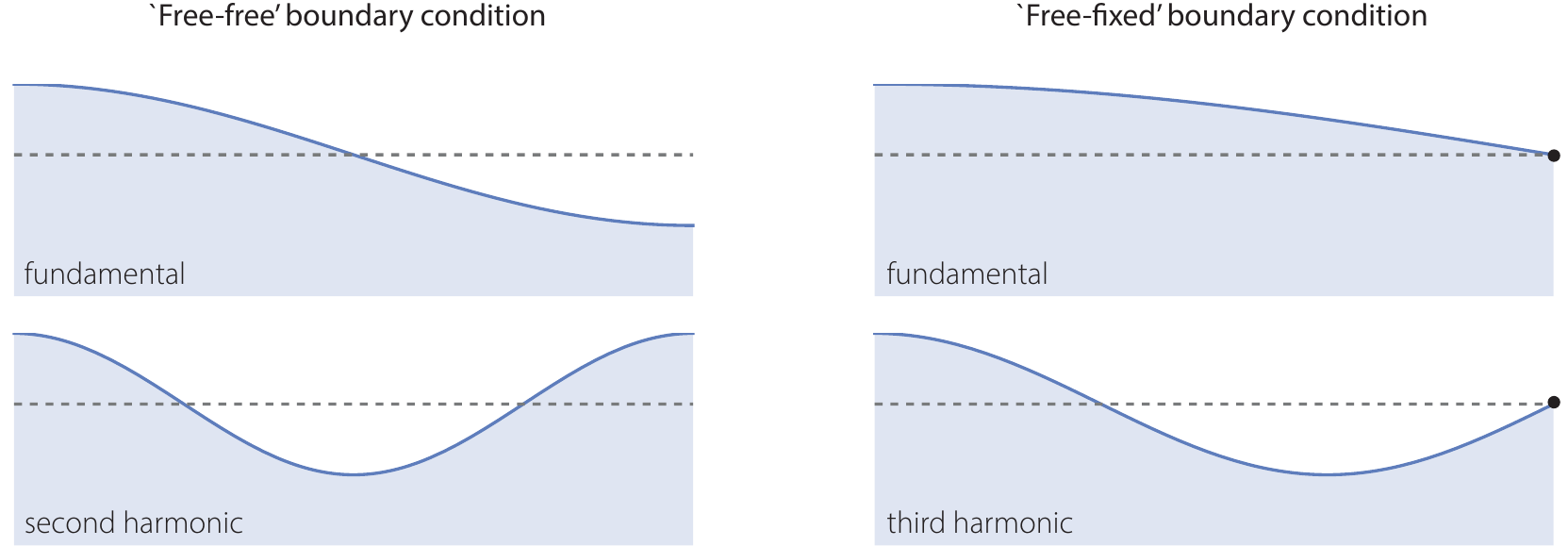}
    \caption{\textbf{Effect of boundary conditions on the harmonic spectrum.} In the case of `free-free' boundary conditions (left), the second resonant mode is at twice the frequency of the fundamental (second harmonic). In the case of `free-fixed' boundary conditions (right), the second resonant mode is at three times the frequency of the fundamental (third harmonic).  }
    \label{fig:boundaryconditionssupp}
\end{figure}

\begin{figure}
    \centering
    \includegraphics[width=\textwidth]{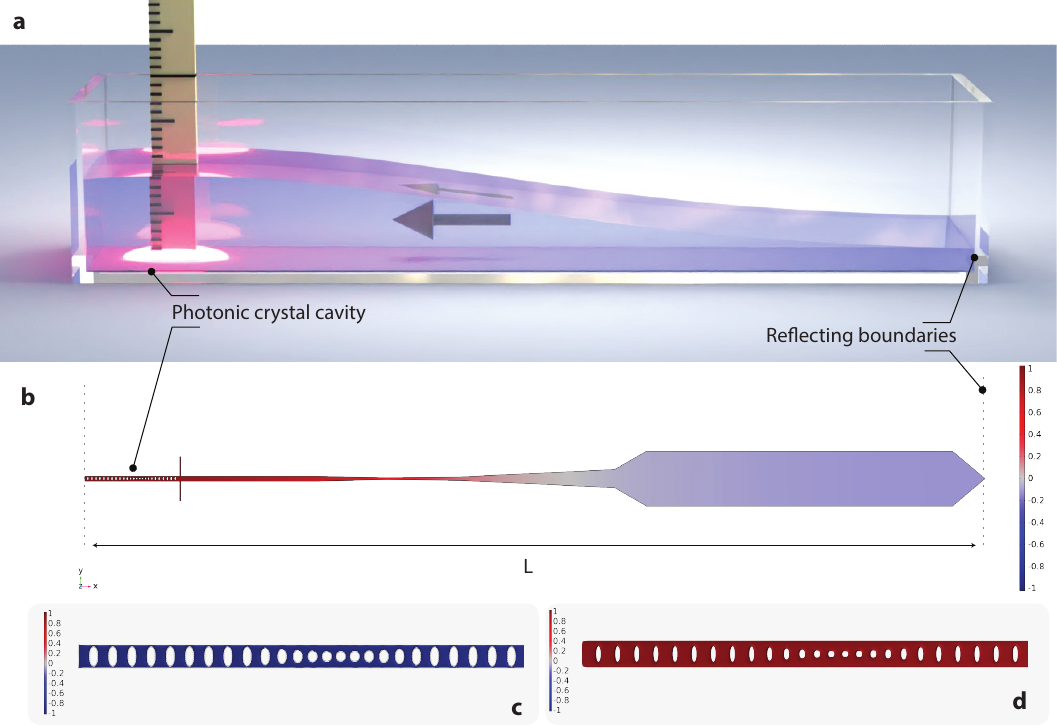}
    \caption{\textbf{Fundamental mode of oscillation of the superfluid wave flume}. a) Artistic illustration of the  fundamental mode of oscillation of the wave tank.  
    b) Finite element method  simulation of the fundamental mode of oscillation of the superfluid wave tank, calculated in the linearized regime. Color scale displays the simulated wave amplitude, normalized to unit displacement $\max\left( H\right)=1$. Displacement amplitude is enhanced at the level of the photonic crystal (left end) compared to the wider support pad (right end)  due to a focusing effect caused by the lateral tapering of the wave flume.  Device length $L=98.5 \mu$m (see Table~\ref{Table_physical_params}).  \textbf{c}  and  \textbf{d}: Zoom-in showing the simulated wave amplitude at the level of the photonic crystal cavity, respectively at phases $\varphi=0$ and $\varphi=180 \degree$, illustrating the mechanism through which the wave motion modulates the cavity resonance frequency.}
    \label{fig:suppFEMacousticmodes}
\end{figure}

Second, the nature of the boundary condition can be independently verified by looking at the device's harmonic spectrum.  This is illustrated in Fig.~\ref{fig:boundaryconditionssupp}. In the case of \emph{`free-free'} boundary conditions (left) the second resonant mode of the wave flume is at twice the frequency of the fundamental mode (corresponding to the second harmonic). In contrast, in the case of \emph{`free-fixed'} boundary conditions (right) the second resonant mode of the wave flume would be at three times the frequency of the fundamental (third harmonic). By varying the frequency $\Omega$ of the drive in the experiments and monitoring the amplitude of the generated waves (see Fig.~\ref{fig:Fig2} of the main text), we can record the mode spectrum of the device (not shown here) and verify that the second resonant mode of the device occurs at close to twice the frequency of the fundamental mode, confirming the \emph{`free-free'} boundary conditions. 

The negligible influence of the glued tapered fiber on the acoustic properties of the device can be understood by considering the large change in acoustic impedance at the contact point between the silicon device and the silica fiber, caused by the rapid change in cross-sectional area. As in the case of a  Helmholtz resonator~\cite{souris_ultralow-dissipation_2017,sawadsky_engineered_2023} this effectively localizes the wave energy to the silicon device. Indeed, as evidenced by the achievable Q factors in excess of $10^2$ (see section~\ref{sectionsuppringdownmeasurements}), only a modest fraction of the wave energy is radiated through the anchoring to the conical tapered fiber. 

\subsubsection{Q factor derived through ringdown measurements}
\label{sectionsuppringdownmeasurements}

\begin{figure}
    \centering
    \includegraphics[width=0.85\linewidth]{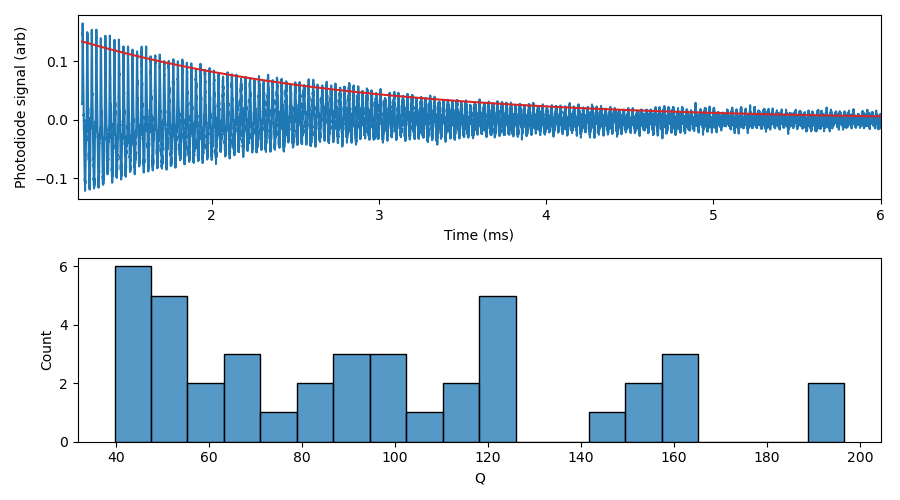}
    \caption{An example ringdown with the fitted exponential envelope, and a distribution of Q factors derived through a series of ringdown measurements.}
    \label{fig:Ringdown_Qextraction}
\end{figure}

The envelope for the ringdown curve is extracted by digitally low-pass filtering the data, and fitting the envelope as an exponential decay. This yields mostly Q factors on the order of 50-200, depending on the experimental settings, with higher laser powers associated with lower acoustic Q factors, as detailed in section~\ref{sectionsuppevaporativedamping}. An example ringdown trace is shown in Fig.~\ref{fig:Ringdown_Qextraction}, with the exponentially decaying amplitude fit shown in red.

We observe that impulse excitations (square impulses) fail to excite a large amplitude waves, and that a large number of sinusoidal excitations are required to efficiently drive the acoustic wave---this is expected from driving a fairly high Q oscillator.

\subsection{Wavelength dependent damping}
\label{sectionatkinsdispersionrelationanddamping}

At non-zero temperatures, third sound should be considered not only as a surface wave which modulates the superfluid depth, but also as a temperature wave. Indeed, wave peaks---where there is an excess of superfluid compared to normal fluid---are colder, while conversely wave troughs are warmer~\cite{atkins_chapter_1970,bergman_third_1971}. (In the zero temperature limit, since the normal fluid density vanishes, this effect is negligible). 
Because of this, fountain pressure and vaporization effects couple to the dynamics of the third sound wave, making for a more complex expression for the phase velocity $v_p$~\cite{atkins_chapter_1970}:
\begin{equation}
    v_p^2=\frac{\Omega^2}{k^2}= \frac{\rho_s}{\rho} g d + \frac{\rho_s}{\rho} \frac{S T\left[ \left( S-\beta/\rho\right) - i \left( K g / \rho \Omega\right) \right]}{ C_{He} - i \left( K L_{He}/\rho \Omega d \right)}.
    \label{eq:vphaseatkins}
\end{equation}
\begin{figure}
    \centering
    \includegraphics[width=.9\textwidth]{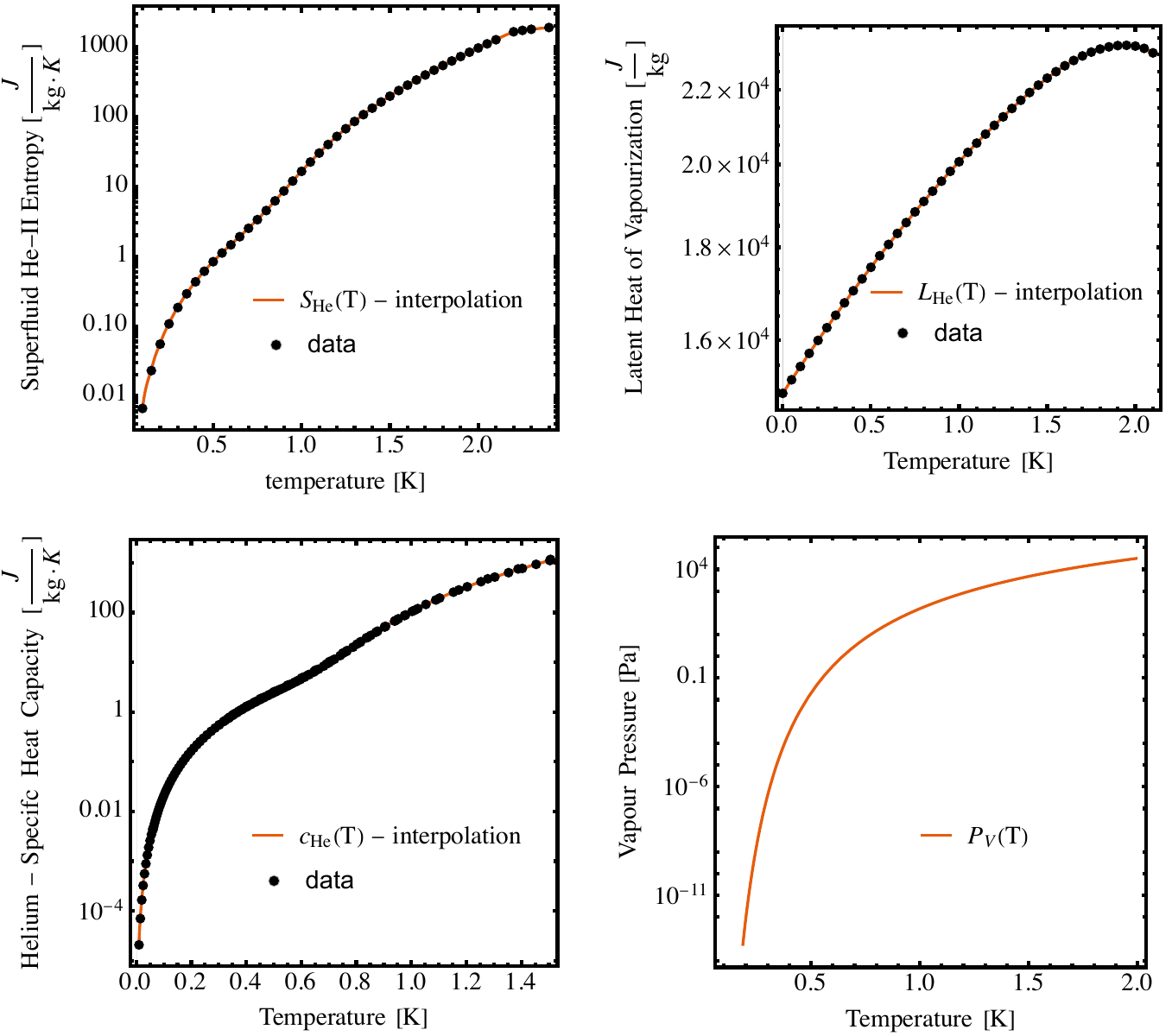}
    \caption{\textbf{Temperature-dependent helium properties}. These are employed in the calculation of the third sound dispersion relation at non-zero temperatures (section~\ref{sectionatkinsdispersionrelationanddamping}) and in the estimation of the evaporative damping of the waves (section \ref{sectionsuppevaporativedamping}).}
    \label{fig:temperaturedependentheliumproperties}
\end{figure}
Here $S$, $C_{He}$ and $L_{He}$ are respectively the temperature-dependent specific entropy of the fluid, its specific heat capacity and the latent heat of vaporization (see Fig.~\ref{fig:temperaturedependentheliumproperties}); $\rho_s$ and $\rho$ are respectively the superfluid density and the total fluid density, and $g$ the acceleration experienced by the helium atoms on the surface of the film, which taking into account both van der Waals and surface tension effects, is
$g=\frac{3 \avdw}{d^4}  +\frac{\sigma}{\rho} k^2$. The function $K$ describes the evaporative process, and takes the form \cite{atkins_chapter_1970}:
\begin{equation}
    K=\gamma \sqrt{\frac{m_{He_{mol}}}{2 \pi R T}} \beta,
\end{equation}
where $R=8.314$ J$\cdot$mol$^{-1} $K$^{-1}$ is the ideal gas constant, $m_{He_{mol}}$ the molar mass of helium, $\gamma$  a coefficient of order unity and $\beta$ is the slope of the vapour pressure curve (see Fig.~\ref{fig:temperaturedependentheliumproperties})~\cite{atkins_chapter_1970}:
\begin{equation}
    \beta=\left(\frac{d P_V}{d T}\right)_{\mathrm{v.p.c.}}
\end{equation}
Equation~\eqref{eq:vphaseatkins} can be numerically solved to yield a dispersion relation of the form $\Omega(k)$. This is now a complex dispersion relation, indicative of losses. 
The addition of evaporative and fountain pressure effects has a negligible effect on wave frequency. This is shown in the left panel of  Fig.~\ref{fig:frequencydependentQthirdsoundatkins}, which plots the full dispersion relation obtained from Eq.~\ref{Eq:qfactoratkins} (blue), along with a dispersion relation accounting only for van der Waals and surface tension restoring forces (dashed orange). Both are overlapped. 
The main contribution of evaporative and fountain pressure effects is to introduce loss. 
This is shown in the right panel of Fig.~\ref{fig:frequencydependentQthirdsoundatkins}, where we plot the wave's quality factor $Q$, defined as:
\begin{equation}
    Q=\frac{\Re(\Omega)}{2 \Im(\Omega)} 
    \label{Eq:qfactoratkins}
\end{equation}
We observe a large decrease in quality factor (increase in loss) for higher wavenumbers (shorter wavelengths), qualitatively consistent with what is observed in the experiments. This explains why the steepness of the wave gradually decays over time, as the higher spatial frequency components are predominantly damped (Fig.~\ref{fig:wavebreakingfulllength}) and  why the generated solitons decay faster than the fundamental acoustic mode of the waveflume (Fig.~\ref{fig:solitonfissionfulllength}).
We note that the inferred  quality factor bottoms out around 500, which is quantitatively  higher than observed in the experiments.  The analysis performed by Atkins~\cite{atkins_chapter_1970} presented here assumes a fluid in thermal equilibrium with its vapor phase ($T_{\mathrm{film}}= T_{\mathrm{vapor}}$). This is not the case here, as the laser power used for actuation and readout means $T_{\mathrm{film}}> T_{\mathrm{vapor}}$. The evaporation rate exceeds the condensation rate, leading to additional loss channels, as discussed in the following section. This effect can be mitigated by operating at lower temperatures, through the use of lower laser powers and a better thermal anchoring of the wave flume to the cryostat.

\begin{figure}
    \centering
    \includegraphics[width=\textwidth]{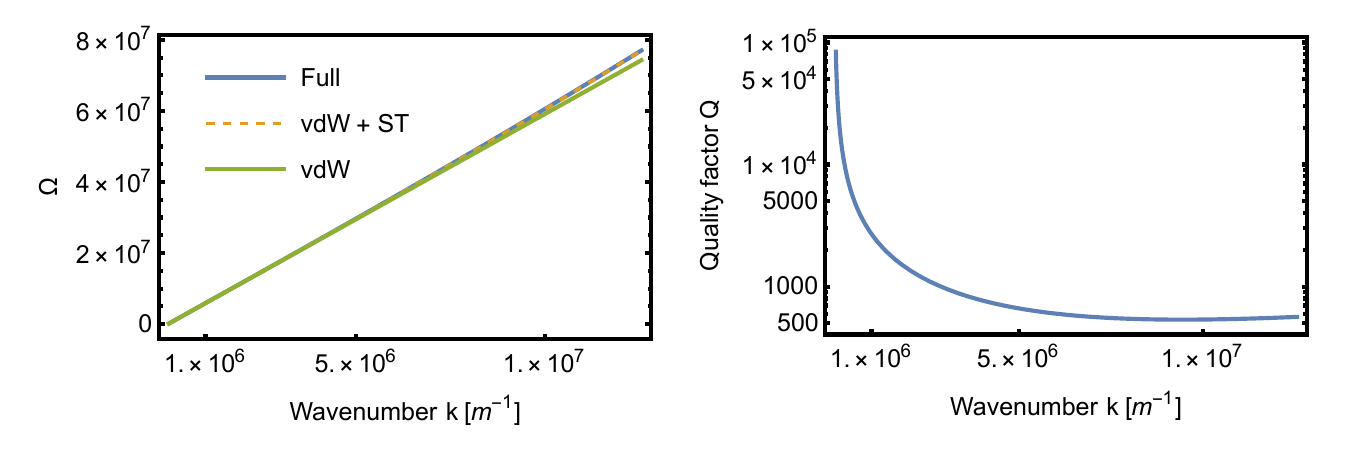}
    \caption{\textbf{Frequency dependent damping of third sound waves}. Left: third sound dispersion relation Eq.~\eqref{eq:vphaseatkins}. Right:  third sound wave quality factor $Q$ calculated  using Eq.~\eqref{Eq:qfactoratkins} (see section~\ref{sectionatkinsdispersionrelationanddamping}). This figure is calculated for a temperature $T=500$ mK, which is a reasonable estimate of the temperature of the superfluid covering the silicon device under microwatt levels of incident optical power (estimated through finite element simulations not shown here).}
    \label{fig:frequencydependentQthirdsoundatkins}
\end{figure}

\subsection{Evaporative damping}
\label{sectionsuppevaporativedamping}

In this section, we further address the effect of laser heating on the acoustic dissipation. While only probed with small optical powers in the nanowatt to microwatt range, the minute device size and only femtoliter volume of helium covering it means evaporation can play a non-negligible role.
Indeed, it should be noted that while the film thickness is kept constant during the measurements---as evidenced by the constant mean surface level in Figs.~\ref{fig:ringdownfulllength}, \ref{fig:wavebreakingfulllength} and \ref{fig:solitonfissionfulllength}---this results from a dynamic equilibrium between a net evaporative flux and ongoing replenishing through film flow.
The magnitude of the evaporative effects can be estimated through energy balance arguments; the evaporated superfluid mass per unit time $\dot{m}_{\mathrm{evap}}$ is given by~\cite{atkins_comments_1954,mcauslan_microphotonic_2016}: 
\begin{equation}
  \dot{m}_{\mathrm{evap}}=\frac{P_d \, \alpha_{abs}}{L_H + T S}.  
\end{equation}
Here $P_d$ and $\alpha_{abs}\in [0,1]$  are respectively the dropped laser power in the optical resonator and  the fraction of dissipated power absorbed as heat, while $T$,  $S$ and $L_H$ correspond respectively to the film temperature, the specific entropy liquid $^4$He and the latent heat of vaporization of superfluid helium~\cite{donnelly_observed_1998} (see Table~\ref{Table_physical_params}). 
In this simplified model, we assume all heat is removed through evaporative cooling by the helium film, and we neglect thermal conduction through the silicon itself. (This is a reasonable assumption given the suspended silicon photonic crystal is only weakly  thermally anchored  to the cryostat via a narrow and long photonic waveguide, itself glued onto a cantilevered tapered silica optical fiber). To estimate $\alpha_{abs}$, we compare the measured optical loss rate in the devices to an estimate of the material limited loss which we obtain by measuring the Q factor of large radius WGM resonators made from the same wafer, and estimate $\alpha_{abs}\simeq 1$ \%. The evaporated mass per oscillation period of the wave $\Delta m$ is given by:
\begin{equation}
   \Delta m=\dot{m}_{\mathrm{evap}}\frac{2\pi}{\Omega}
\end{equation}
This value can be related to the superfluid mass initially present on the device $m_{\mathrm{He}}=6\times 10^{-16}$ kg (see Table~\ref{Table_physical_params}).  For traces \ref{fig:ringdownfulllength}, \ref{fig:wavebreakingfulllength} and \ref{fig:solitonfissionfulllength}, we find that the ratio $\Delta m/m_{\mathrm{He}}$ giving the fraction of the mass in the superfluid wave flume evaporated every oscillation period, is respectively 0.42\%; 1.7\% and 3.4 \%. Assuming the evaporated atoms carry on average the same proportion of the wave energy as those remaining in the tank, this would impose an upper bound on the acoustic quality factor of 240; 60 and 30 respectively, which is broadly consistent with the observed values, indicating that evaporative damping plays an important role at higher optical powers.

This platform presents the opportunity for an in-depth investigation of evaporative damping of fluidic oscillators. This includes studying the additional damping contribution due to the stochastic recoil from evaporating helium atoms not considered here. This will be pursued in further work.

\section{Hydrodynamic Modeling}
\label{methodssectionhydrodynamicmodelling}

We support our experimental observations with hydrodynamical simulations of the third sound waves as detailed in this section.  In the experiment, the temperature is sufficiently low that the superfluid fraction is approximately 100\%, and hence the influence of the viscous normal fluid component may  be neglected~\cite{nakajima_solitary_1980}. The inviscid superflow is therefore potential and incompressible everywhere in the fluid, describing perfect potential flow, i.e., 
\begin{equation}
\nabla^2 \phi = 0,
\end{equation}
and the equations of motion on the surface of the fluid, defined by $y = \eta(x,t)$ are ~\cite{nakajima_solitary_1980}
\begin{gather}
\eta_t +  \phi_x \eta_x  = \phi_y, \label{eqn:eta1} \\
\phi_t  -\alpha\left[\frac{1}{(h+\eta)^3} - \frac{1}{h^3} \right] + \frac{1}{2} (\phi_x^2 + \phi_y^2) = 0. \label{eqn:phi1}
\end{gather}
 \begin{figure}
\includegraphics[width=\textwidth]{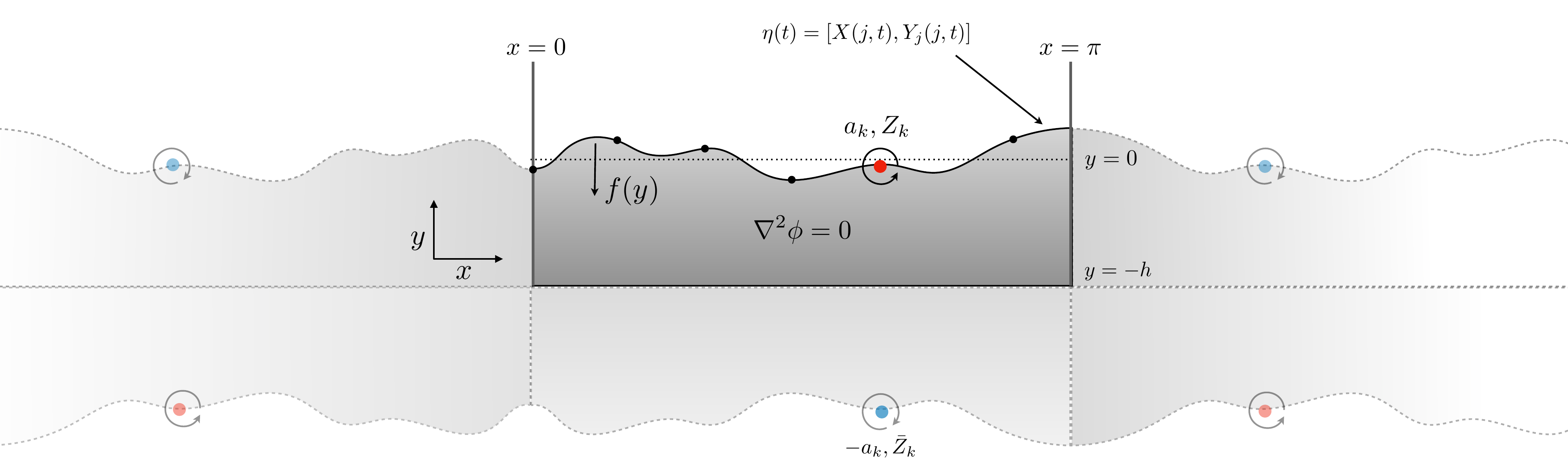}
 \caption{Schematic showing the numerical setup of the problem. The interface between two fluids can be modelled as a distribution of singularities along the surface $S(t)$, i.e., a discretised vortex sheet comprised of point vortices with strengths $a_k$ located at positions $Z_k = X_k +iY_k$. Hard wall boundaries at $x=0,\pi$ are implemented by using periodic boundary conditions on the domain $x \in [0,2\pi]$, with image vortices constrained by reflection symmetry about $x=\pi$. Similarly, the hard wall bottom at $y=-h$ is incorporated by the use of image vortices of strengths $-a_k$ located at $\bar{Z_k} = X_k - iY_k$.}
 \label{fig:sloshing}
 \end{figure}
Here $\alpha$ is the van der Waals coefficient, $h$ is the ambient thickness of the film, and $\phi_x \equiv u$ and $\phi_y \equiv v$ are the velocities along $x$ and $y$, respectively. For brevity, we have used notation where subscripts denote the derivatives. The equations are the same as those that govern gravity waves on the surface of water, except that the restoring force is nonlinear; for small $\eta$, one may expand the van der Waals term to first order in $\eta$, producing an effective gravitational term $f(\eta) \approx g\eta + \mathcal{O} (\eta^2)$.

For boundary conditions, we model the superfluid resonator as a 2D rectangular container of length $L$ with vertical walls at $x = \pm L/2$. The potential $\phi$ and the film surface height $\eta$ thus satisfy the spatial boundary conditions
\begin{align}
\partial_x \phi &= 0, & (&x = \pm L/2); \\
\partial_y \phi &= 0, & (&y = -h); \\
\partial_x \eta &=0, & (&x = \pm L/2). 
\end{align}
The first two equations state that the fluid cannot flow through the boundary, while the third imposes a free boundary condition at the edge. 

Accurate modeling of the experiment poses a computational challenge due to the extreme aspect ratio ($h/L\ll 1$), the large amplitude waves excited ($H/h \sim 1$) and the nonlinear restoring force arising from the van der Waals potential. 
Given these considerations, we opt for a scheme which solves the full hydrodynamic equations (\ref{eqn:eta1}) and (\ref{eqn:phi1}), rather than making shallow-water-type approximations. This requires a robust and accurate computational scheme. We therefore adopt the spectrally accurate semi-Lagrangian boundary integral approach developed by Roberts, originally in the context of extreme water waves. Full details of the derivation of the method can be found in Refs~\cite{roberts1983stable,smith1999branching}.

The essence of the formulation is to model the fluid interface as a vortex sheet, which may be discretized as a set of singularities (point vortices)  along the surface [see Fig.~\ref{fig:sloshing}]. The strengths (circulations) of the singularities depend on the instantaneous distribution of the surface profile. The interface $\eta(x,t)$ is parameterized using an auxiliary variable $j$, giving a discrete set of points $X_j(t), Y_j(t)$, with $j = 0,\dots N-1$. The points $X_j,Y_j$ follow the fluid flow at the interface and hence evolve according to
\begin{align}
\frac{\partial X_j}{\partial t} &= u_j,  &  \frac{\partial Y_j}{\partial t} & = v_j, 
\label{eqn:tracerEquation}
\end{align}
where $u_j$ and $v_j$ are the velocities of the fluid interface, at the point $(X_j,Y_j)$, in the $x$ and $y$ directions respectively.   For our numerical analysis, it is convenient to non-dimensionalize the equations by working in units of length and time as
$x_0 = L/\pi$  and $t_0 = \left[ g \pi/L \right]^{-1/2}$ respectively. In these units the container length is $\pi$. Working in these units, in terms of the Lagrangian coordinates,  Eq.~(\ref{eqn:phi1}) becomes ~\cite{roberts1983stable}
\begin{equation}
\frac{\partial \phi_j}{\partial t}  = \frac{h}{3}\left[ \frac{1}{(1 + Y_j/h)^3} - 1 \right] + \frac{1}{2} \left( u_j^2 + v_j^2 \right),
\label{eqn:phiEquation}
\end{equation}
Note the change in sign of the last term (in comparison with Eq.~(\ref{eqn:phi1})), due to the change to the Lagrangian description.  We enforce the system is periodic in $j$ with period $N$, and so the Dirichlet boundary conditions at $x = 0,\pi$ are implemented by setting the values of $\{X_j,Y_j,\phi_j\}$ for $j = N/2+1,\dots,N-1$ to be the mirror reflection of the points $j=1,\dots,N/2-1$ about the plane $x=\pi$. Additionally, $u_j=0$ for the points $j=\{0,N/2\}$ , which must be constrained to lie at the walls $x=\{0,\pi\}$ throughout the evolution.

The dynamics can then be calculated as follows. Defining $Z_k = X_k + i Y_k$, and the vortex strengths $a_k$, the potential and vortex strengths are related via the relation
\begin{align}
\phi_k'  &= \left[ \frac{1}{2}  + \frac{1}{4 \pi} \Im \left( \tfrac{Z_k''}{Z_k'} \right) \right]a_k    +  \frac{1}{4 \pi} \sum_{\substack{j=0 \\ j \neq k}}^{N-1} a_j \Im \left[ Z_k' \cot \left( \tfrac{Z_k - Z_j}{2}\right) \right]   -  \frac{1}{4 \pi} \sum_{j=0}^{N-1}  a_j \Im \left[\cot \left( \tfrac{Z_k - \bar{Z}_j + 2 i h}{2}\right) \right]
\label{eqn:linearSystem}
\end{align}
where $\Im$ denotes the imaginary part, and the overbar denotes the complex conjugate.  Here, the second term represents the interaction between different vortices on the surface, and the third term represents the interaction with the images reflected about $y=0$. The first term arises from regularizing the singular term $j=k$. The cotangent terms arise from considering the usual logarithmic potential of point vortices, and summing over the infinite series of images periodic in $x$. The primes denote differentiation with respect to the auxiliary variable $k$. These are evaluated using Fourier transforms, taking care to correctly zero the sawtooth mode~\cite{roberts1983stable}. 

Eq.~(\ref{eqn:linearSystem}) gives a diagonally-dominant linear system of the form $M_{kj} a_j = \phi'_k$ that can be solved for the vortex strengths. From the vortex strengths, the velocities are then obtained from the complex velocity $w_k = u_j + iv_k$, via the relation
\begin{align}
\bar{w}_k = \frac{a_k}{2 Z'_k}  - \frac{i}{4\pi} \left( \frac{Z''_k a_k}{(Z'_k) ^2}   - \frac{2 a'_k}{ Z'_k} \right)  - \frac{i}{4\pi} \sum_{\substack{j=0  j \neq k}}^{N-1} a_j \cot \left( \tfrac{Z_k - Z_j}{2} \right)   + \frac{i}{4\pi} \sum_{j=0}^{N-1} \cot\left( \tfrac{Z_k - \bar{Z}_j + 2ih}{2} \right).
\label{eqn:complexVelocity}
\end{align}
This completes the procedure to evolve the dynamics in time; first, given an initial surface profile $Z_k$ and potential $\phi_k$, we calculate the vortex strengths by solving the linear system Eq.~(\ref{eqn:linearSystem}). We then calculate  the velocities from Eq.~(\ref{eqn:complexVelocity}). The positions and potential are then updated according to Eqs.~(\ref{eqn:tracerEquation}) and~(\ref{eqn:phiEquation}). Throughout we use a 4th order Runge-Kutta scheme to do the timestepping.

Due to the extreme conditions and long simulation times required, to ensure accuracy of the solutions, we verify that the relevant conserved quantities are conserved to within an acceptable tolerance; in particular the Hamiltonian which describes the total energy of the wave system is 
\begin{equation}
H =   \int \mathrm{d}x\; [T(x) + V(x)]
\end{equation}
with
\begin{equation}
T(x) =   \frac{1}{2}\int_{-h}^{\eta}  |\nabla \phi|^2 \mathrm{d}y, \; 
\end{equation}
and 
\begin{equation}
V(x) =  \frac{\alpha}{2} \left[\frac{1}{(h + \eta)^2} - \frac{1}{h^2} \right] ,
\end{equation}
being the kinetic and potential energies of the surface waves respectively. In addition, conservation of mass requires that the mean surface height satisfies the condition
\begin{equation}
\bar{\eta} = \int \mathrm{d}x \; \eta(x,t) = 0. 
\end{equation}
Details of calculating these quantities are provided in Ref.~\cite{roberts1983stable}.
Accurate conservation of these quantities requires that the number of collocation points $N$ used to generate a solution must satisfy $L/N \ll h$; in practice, for our most extreme aspect ratio $h/L \sim 300$, this requires $N\sim 2048$ collocation points.

For the simulations shown in the main text and in the Supplementary video S1, a sinusoidal wave with  amplitude $H$ equal to 10\% of the fluid's depth $h$ is initialized,  and evolved under the scheme described above for 50 periods, which takes approximately 4 days on a desktop PC. The simulated wave amplitude $H/h$ is smaller than that achieved in the experiments; we choose to initialize a smaller amplitude wave---and instead let it evolve over more wave cycles---as the simulations become numerically unstable at larger amplitudes $H/h>0.2$. The amplitude plotted is taken from a location 4\% of the resonator's length away from the boundary (red marker in the top panel of movie S1), and is spatially averaged over a width of 3\% of the resonators length, to correspond to the location and spatial extent of the photonic crystal mode.  
The simulated wave flume has an aspect ratio of $h/L=300:1$, with 2048 points tracking the surface of the wave. The mean surface height is conserved to within $\sim 10^{-8}$, and the energy is conserved to within $0.02\%$ of its initial value (see movie S1, bottom panel). As shown in supplementary movie S1, the sinusoidal wave is evolved for 27 periods before matching the profile of multisoliton fission observed in the experiments  (Fig.4 of the main text).

\section{Wave dispersion}
\label{sectionappendixdispersion}

In this Appendix, we perform a detailed analysis of the wave dispersion in our superfluid wave flume. A proper understanding of the both the magnitude and the sign of the dispersion is crucial to analyze our experimental results. Indeed, as  discussed in the following and \ref{sectionappendixinfluenceofdispersionandnonlinearity}, in conjunction with the size and magnitude of the nonlinearity, the magnitude of the dispersion ultimately determines the number of generated solitons, while the sign of the dispersion determines the `bright' vs `dark' nature of the soliton solutions (\textit{i.e.} whether they present as a traveling elevation above the mean fluid baseline, or a depression below the baseline respectively).

As detailed in Ref.~\cite{trillo_experimental_2016}, the number of solitons emerging from the fission process $N_s$ scales  
as $1/\epsilon$, where the dimensionless parameter $\epsilon^2$ is defined as a ratio of the nonlinear length scale $L_{nl}$ over which the wave critically steepens
to the dispersive  length scale $L_d$:
\begin{equation}
    \epsilon^2=6 L_{nl}/L_d 
    \label{Eqepsilon}
\end{equation}
and is inversely proportional to the Ursell number.
For our system, considering only the van der Waals nonlinearity and the small hydrodynamic dispersion in the shallow water limit, Eq.~\eqref{Eqepsilon}  predicts a very large number of solitons, of the order of several hundred, whereas the observed number is more than one order of magnitude smaller.  
This discrepancy is indicative of a much larger dispersion at play here.

We therefore perform a detailed analysis of many potential sources of dispersion in our experiment.
The magnitude and sign of these are briefly addressed here, with full details in the following sub-sections. 
First, shallow fluid waves are naturally (very weakly) dispersive, with shorter wavelengths (higher wavenumbers $k$) propagating with a lower group velocity. This corresponds to a weak so-called \emph{normal} dispersion.

Surface tension is also present in superfluid helium films. The effect of surface tension is in contrast to increase the group velocity of shorter wavelength waves.  This introduces a comparatively stronger \emph{anomalous} dispersion, and is discussed in section~\ref{sectionsuppcapillarydispersion}
. We additionally consider the fact that the waves do not propagate inside a 1-dimensional wave flume, but rather along the outer surface of a cylinder-like object (see Fig.~\ref{fig:suppFEMacousticmodes}). The effect of surface tension in this case is simply to provide a correction to the effective van der Waals acceleration felt by the film (see section~\ref{subsectioninfluenceofsurfacecurvature}).

Next, we consider the effect of operating at finite temperatures. At non-zero temperatures, the third sound wave takes on a thermal wave nature as well, leading to a modified dispersion relation due to fountain pressure restoring forces. This so-called thermomechanical contribution to the dispersion~\cite{atkins_chapter_1970, bergman_third_1971} is evaluated in section~\ref{sectionatkinsdispersionrelationanddamping} and is negligible here.
Finally, we consider the dispersion introduced by the photonic crystal cavity at the end of the wave flume. As detailed in section~\ref{sectionsuppacousticbraggmirrordispersion}, this periodic hole pattern also forms a \emph{phononic} band-gap, with its associated strong dispersion at the band edge. We find that this introduces a strong normal dispersion, which eclipses the anomalous dispersion caused by surface tension. 

Taking all these sources of dispersion into account, the superfluid waves experience a normal dispersion with a magnitude $\sim 35$ times larger than expected from the hydrodynamical dispersion  alone. Because of this, the behaviour of the wave flume can be well emulated with a hydrodynamic simulation (see section~\ref{methodssectionhydrodynamicmodelling}), without the need for surface tension, whose aspect ratio has been reduced approximately $\sim 35$ fold to  $300:1$, as can be seen in the good experimental agreement in Figs. 4 and 5 of the main text. (Reducing the aspect ratio by a factor $x$ increases the wavenumber of the fundamental mode by $x$, and increases the magnitude of the dispersion by $~x$ in the shallow fluid limit, see \eqref{Eq:dispersionrelationnosurfacetension}).


\subsection{Capillary dispersion}
\label{sectionsuppcapillarydispersion}

Here we quantify the influence of surface tension on third sound dispersion. Third sound waves in the absence of surface tension have the following dispersion relation:
\begin{equation}
    \Omega_{no ST}\left(k, d\right) = \sqrt{g_{vdW}\,k \tanh(k\, d)}=\sqrt{ \frac{3 \avdw}{d^4} k   \tanh(k\, d)},
    \label{Eq:dispersionrelationnosurfacetension}
\end{equation}
where $k=2\pi/\lambda$ is the wavenumber and, in the low temperature limit, the acceleration is given by the van der Waals attraction to the substrate $g_{\mathrm{vdw}}=\frac{3 \avdw}{d^4}$.
In the presence of surface tension, the effective acceleration is modified, and the dispersion relation takes the form:
\begin{equation}
    \Omega\left(k, d\right) = \sqrt{\left( \frac{3 \avdw}{d^4}  +\frac{\sigma}{\rho} k^2 \right) k \tanh(k\, d)},
    \label{Eq:dispersionrelationsurfacetension}
\end{equation}
where $\sigma$ and $\rho$ are respectively the surface tension and density of superfluid $^4$He, see Table~\ref{Table_physical_params}. The magnitude of van der Waals and capillary accelerations become identical for a wavelength $\lambda=2\pi\sqrt{\frac{\sigma}{\rho g_{\mathrm{vdw}}}}$, equal to 140 nm for the 6.7 nm film thickness considered here---with longer wavelengths dominated by the van der Waals interaction and shorter wavelengths by surface tension.  For the fundamental wavelength of the flume ($\lambda=197$ $\mu$m), the influence of surface tension on the wave velocity is negligible ($\Omega/\Omega_{no ST}=1 + \mathcal{O}(10^{-7})$), however its role in the dispersion must be accounted for, as the hydrodynamic dispersion (Eq.~\ref{Eq:dispersionrelationnosurfacetension}) is itself quite small in the shallow water limit $\lambda\gg d$.

Similarly, we define the dispersion relation for purely capillary waves:
\begin{equation}
    \Omega_{cap}\left(k, d\right) = \sqrt{  \frac{\sigma}{\rho} k^3 \tanh(k\, d)},
    \label{Eq:dispersionrelationcapillarywave}
\end{equation}

\begin{figure}
    \centering
    \includegraphics[width=\textwidth]{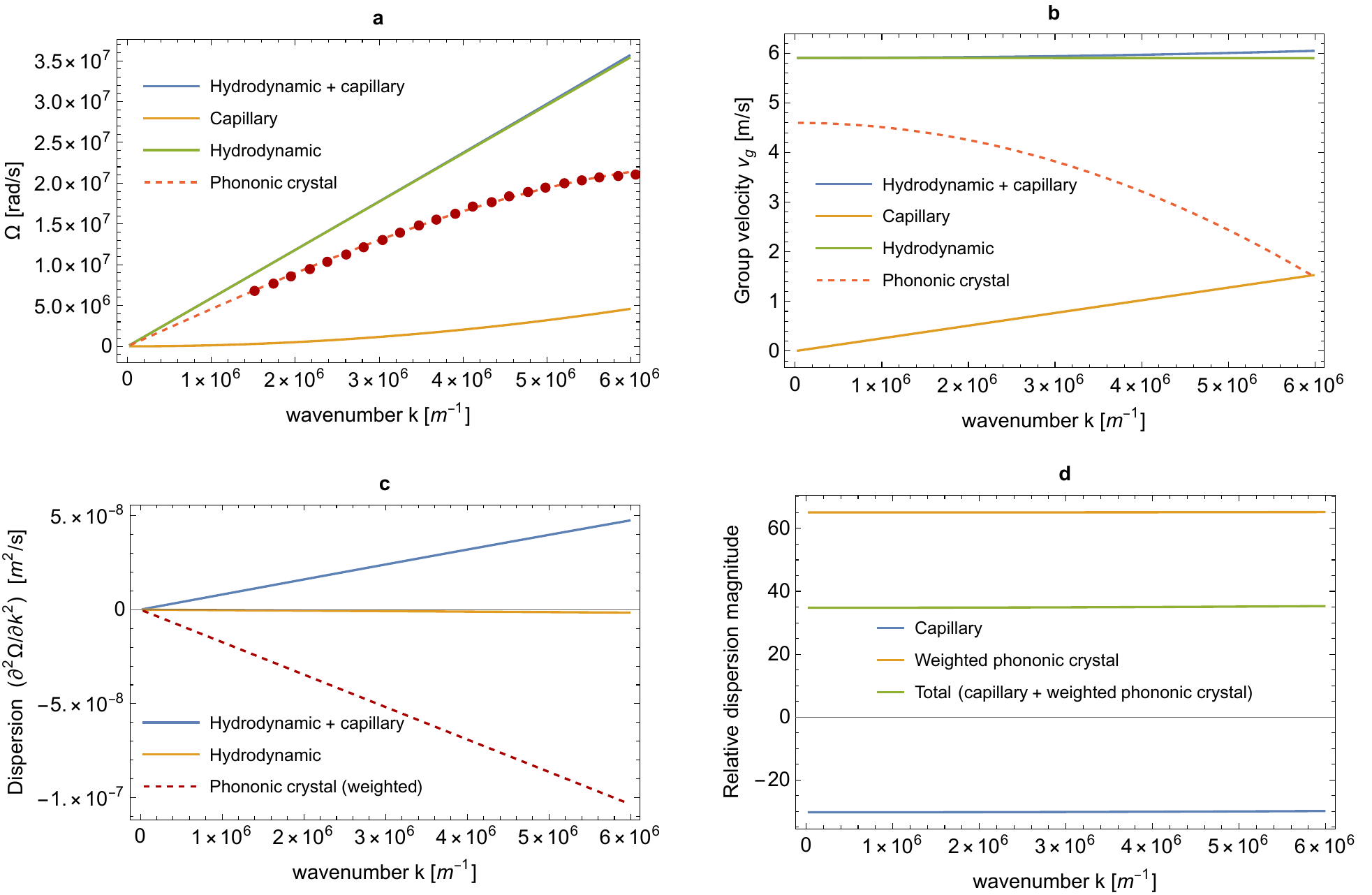}
    \caption{\textbf{Analysis of dispersion mechanisms in the superfluid wave flume.} \textbf{a} Green: hydrodynamic dispersion relation, obtained from  Eq.~\eqref{Eq:dispersionrelationnosurfacetension}; Blue: hydrodynamic dispersion relation taking into account capillary effects, obtained from~Eq.~\eqref{Eq:dispersionrelationsurfacetension}. Orange: capillary dispersion relation, obtained from Eq.~\eqref{Eq:dispersionrelationcapillarywave}. Dashed red:  Phononic crystal dispersion relation, see Fig.~\ref{fig:bandgap} and section~\ref{sectionsuppacousticbraggmirrordispersion} for more details.
    \textbf{b} Group velocity $v_g=\left(\frac{\partial \Omega}{\partial k}\right)$ associated with the different dispersion relations plotted in \textbf{a}. Green: hydrodynamic case.
    Blue: hydrodynamic and capillary contributions. Orange: purely capillary case.  Dashed red: group velocity for an acoustic wave traveling through the phononic crystal.
    \textbf{c} Dispersion $\left(\frac{\partial^2 \Omega}{\partial k^2}\right)=\frac{\partial v_g}{\partial k}$, obtained from \textbf{b}. Blue: ; Orange:  ;  Dashed red:   .   \textbf{d} Relative dispersion magnitude   Blue: ratio of capillary to hydrodynamic dispersion magnitude. Orange:  Ratio of weighted phononic crystal dispersion to hydrodynamic dispersion.   Green: Ratio of total (capillary + weighted phononic crystal) dispersion to hydrodynamic dispersion.}
    \label{fig:dispersioncomparison}
\end{figure}

We find that the inclusion of surface tension increases the magnitude of the dispersion approximately thirty fold (see Fig.~\ref{fig:dispersioncomparison} and following section~\ref{sectionsuppanalysisofcombinedsourcesofdispersion}). The magnitude of this increase is in line with what is expected. However the sign of this dispersion (anomalous, with high wavenumbers propagating with increased group velocity) is opposite to the normal dispersion  observed in the experiments. Indeed, as detailed in section~\ref{sectionappendixinfluenceofdispersionandnonlinearity}, this combination of nonlinearity and dispersion would generate cold solitons, presenting as a traveling elevation above the baseline, as predicted by Nakajima et al.~\cite{nakajima_solitary_1980}.

\subsection{Acoustic Bragg grating dispersion}
\label{sectionsuppacousticbraggmirrordispersion}

\begin{figure}
    \centering
    \includegraphics[width=\textwidth]{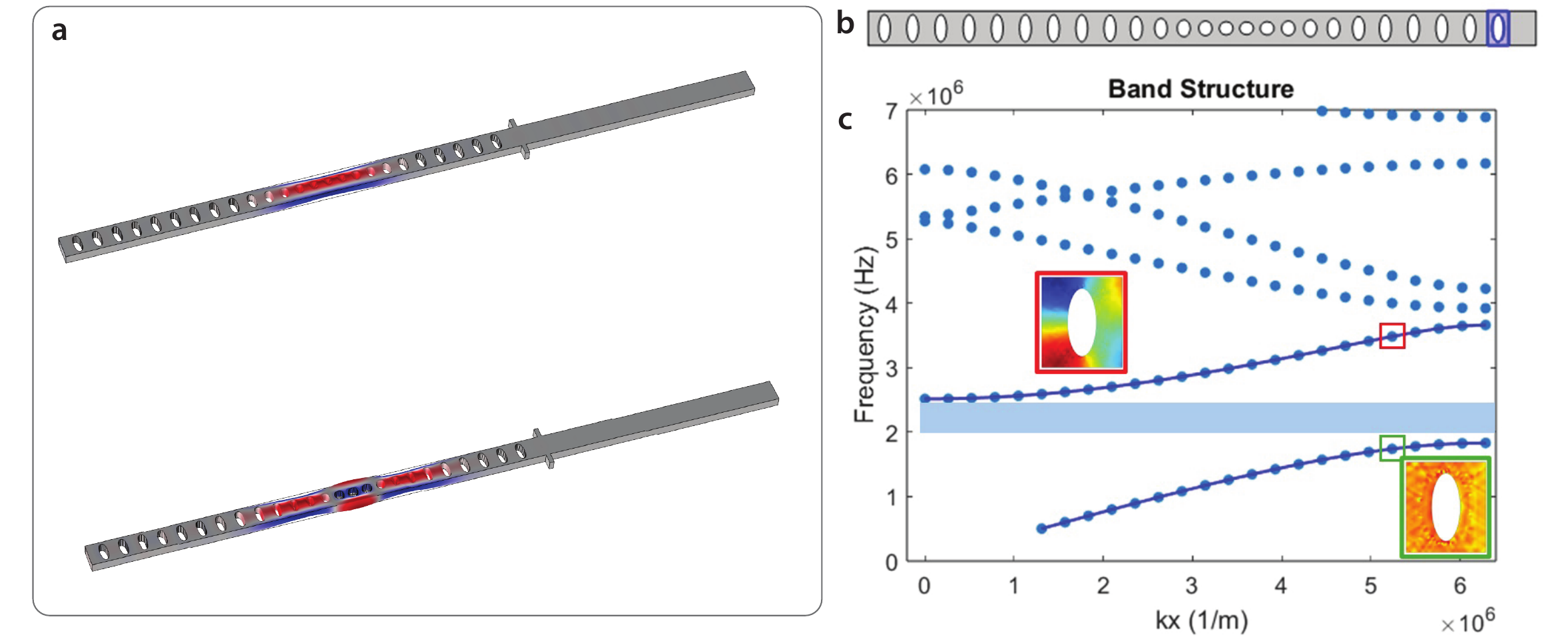}
    \caption{\textbf{Superfluid phononic crystal.} \textbf{a} Examples of superfluid acoustic modes localized within the acoustic bandgap by the tapering of the central crystal holes. With $d_0=6.7$ nm, top mode frequency $\Omega/2\pi=6.9$ MHz; bottom mode frequency $\Omega/2\pi=7.3$ MHz. \textbf{b} Top view of the photonic/phononic crystal cavity. The blue rectangle marks the unit cell used for the band structure calculation. \textbf{c} Superfluid acoustic band structure.  The acoustics for the photonic crystal cavity were simulated in COMSOL Multiphysics for a unit cell of dimensions $500\times 560$ nm (see \textbf{b}) using the \textit{Linearized Euler, Frequency Domain} module~\cite{forstner_modelling_2019}, where Floquet boundary conditions were defined along the direction of periodicity. The band structure was created by plotting the simulated frequency for different values of the Floquet vector $k_x$.}
    \label{fig:bandgap}
\end{figure}

Here we calculate the dispersion caused by the acoustic Bragg grating 
formed by the photonic and phononic crystal at the end of the wave flume. Indeed, the photonic crystal holes at the end of the wave flume (see Figs.~\ref{fig:supplementaryfabrication}, \ref{fig:suppFEMacousticmodes} and \ref{fig:photoniccrystalopticalmodes}) also result in a periodic modulation of the acoustic impedance, and thus forms both a \emph{photonic} and \emph{phononic} resonator~\cite{sfendla_extreme_2021, korsch_phononic_2024}. Figure~\ref{fig:bandgap} shows the calculated acoustic bandgap, following the method outlined in Refs.~\cite{sfendla_extreme_2021,forstner_modelling_2019}, alongside  examples of acoustic modes confined within the acoustic bandgap due to the tapering of the central holes, calculated using the method outlined in section~\ref{methodsectionacousticeigenmodes}. The dispersive effects can be well seen in Fig.~\ref{fig:bandgap}(c), through the changing slope of the dispersion curve ($v_g=\frac{\partial \Omega}{\partial k}$) near the edge of the bandgap.
This calculation provides the group velocity of a wave traveling through an infinite periodic medium, with the unit cell highlighted in blue in Fig.~\ref{fig:bandgap}(b), while the superfluid wave interacts with the Bragg mirror only twice per round-trip.
To account for this, we introduce a weighting factor corresponding to the spatial extent of the phononic crystal divided by the total length of the wave flume, which corresponds to  10\%.
The dispersive effects of Bragg mirrors near the edge of the stop band are well known and widely used in the photonic realm for dispersion compensation, slow light generation and sensing~\cite{skolianos_slow_2016}, and with chirped gratings employed for dispersion control~\cite{ouellette_dispersion_1987}. 
Such techniques could be leveraged in future work as an additional control parameter, as has been demonstrated in acoustic waveguides~\cite{kurosu_-chip_2018}.

\subsection{Analysis of combined sources of dispersion}
\label{sectionsuppanalysisofcombinedsourcesofdispersion}

We plot these combined dispersion relations in Fig.~\ref{fig:dispersioncomparison}\textbf{a}, for the range of wavenumbers relevant in our experiment. (The maximum value of $k$ corresponds to a wavelength of 1 $\mu$m, i.e. 200 times shorter than the fundamental wavelength of oscillation of the flume). The green trace plots the hydrodynamic dispersion relation for shallow fluid waves, obtained from  Eq.~\eqref{Eq:dispersionrelationnosurfacetension}. It is nearly linear, as expected from the weakly dispersive nature of waves in our shallow limit $\lambda \gg h$. The blue trace (nearly overlapped with the green trace) shows the correction introduced by the addition of surface tension (Eq.~\eqref{Eq:dispersionrelationsurfacetension}). Its effect on the frequency is minimal, as we consider waves whose wavelength is still well in excess of the capillary length. The orange trace plots the dispersion relation for purely capillary waves (Eq.~\eqref{Eq:dispersionrelationcapillarywave}).
The solid red dots correspond to the calculated dispersion relation for a third sound wave traveling through a phononic crystal like the one present at the end of the wave flume. These are obtained through finite element simulations, as shown in Fig.~\ref{fig:bandgap}. The dashed red line corresponds to an analytical interpolation of the data points.

The group velocity $v_g$ associated with these dispersion relations is plotted in Fig.~\ref{fig:dispersioncomparison}\textbf{b}. The group velocity associated with Eq.~\eqref{Eq:dispersionrelationnosurfacetension} is nearly constant over the wavenumber range of interest, and close to the dispersion-less velocity $c_3=\sqrt{g_{vdw} \,h}=\sqrt{\frac{3 \avdw}{h^3}}=5.9$ m/s.
The addition of capillary restoring forces increases the group velocity by a modest amount at higher wavenumbers (blue trace). In contrast, the group velocity of a third sound wave traveling through a phononic crystal as the one present at the end of the wave flume is significantly reduced, particularly as the bandgap is approached (dashed red line). 

The magnitude of the dispersion (defined here as $\left(\frac{\partial^2 \Omega}{\partial k^2}\right)$\footnote{Related to the group velocity dispersion $\left(\frac{\partial^2 k}{\partial \Omega^2}\right)$ by a factor $\frac{-1}{v_g^3}$.} is plotted in Fig.~\ref{fig:dispersioncomparison}. The orange trace plots the dispersion due to the shallow wave equation alone (Eq.~\eqref{Eq:dispersionrelationnosurfacetension}). The blue trace corresponds to the dispersion due to the inclusion of capillary effects. The dashed red trace plots the dispersion caused by the phononic crystal. It has been weighted by a factor of 10\%, to account for the phononic crystal covering only 10\% of the length of the flume. 
Finally, Fig.~\ref{fig:dispersioncomparison}\textbf{d} plots the relative magnitude of the dispersion mechanisms, compared to the hydrodynamical dispersion alone. The blue trace plots the ratio of capillary to hydrodynamic dispersions. Its value of $-30$ means the dispersion caused by capillary effects is here 30 times larger in magnitude, but of opposite sign (anomalous vs normal) compared to the hydrodynamic dispersion. Similarly, the dispersion due to the phononic crystal is approximately 65 times larger than the hydrodynamic dispersion, for the range of wavenumbers relevant in our experiment. The total contribution of capillary and phononic crystal is therefore a dispersion which is 35 times larger (and of same sign) compared to that expected from the shallow fluid wave dispersion relation alone. 
As mentioned earlier, in the extremely shallow limit $\lambda\gg h$ considered here, the magnitude of the hydrodynamic dispersion is inversely proportional to the wavelength. This means that the effectively 35 times larger dispersion can be well emulated with a hydrodynamic simulation---which does not require surface tension---whose aspect ratio has been reduced approximately $\sim 35$ fold to  $300:1$, as can be seen in the good experimental agreement in Figs. 4 and 5 of the main text.

\begin{table}
    \centering
    \begin{tabular}{l c c c c }
    \hline
      Parameter   &  Symbol & Value & Unit & Source\\
      \hline
      \textbf{Optical resonator}\\
      \hline
    Intrinsic energy decay rate & $\kappa_i/2\pi$ & 80 & GHz & meas. \\
    Extrinsic energy decay rate & $\kappaex/2\pi$ & 27 & GHz & meas. \\
    Absorptive decay rate & $\kappa_{\mathrm{abs}}$ & 0.67 & GHz  & \\
    Optomechanical coupling & $G/2\pi$ & $5.6$ & GHz/nm & FEM\\
    Resonance wavelength & $\lambda_0$ & 1598 & nm & meas. \\
    \hline
      \textbf{Superfluid third sound mode}\\
      Frequency & $\Omega_M/2\pi$ & $27$ & kHz & meas.\\
      Wavelength fundamental mode & $\lambda$ & $2L=197$ & $\mu$m & meas. \\
      Decay rate & $\Gamma$ & $2\pi\times 270$ & Hz & \\
     Mean superfluid film thickness & $d_0$ & 6.7 & nm & meas.\\ 
     Third sound speed & $c_3$ &$5.9$ & m/s\\
      Van der Waals coefficient for silicon & $\avdw$ & $3.5\times 10^{-24}$& m$^5$s$^{-2}$ & \cite{baker_theoretical_2016, sabisky_onset_1973}\\
      \hline
     \textbf{Device geometry}\\
      Resonator length & $L$ & 98.5 &$\mu$m & SEM\\
       Silicon resonator surface area & $\mathcal{A}$ & $5\times10^{-10}$ & m$^2$ & SEM\\
      Superfluid $^4$He density   & $\rho_{\mathrm{He}}$ & 145 & kg/m$^3$ & \cite{donnelly_observed_1998}\\
     Superfluid helium volume & $V_{\mathrm{He}}=A\times d_0$ &  $ 4\times 10^{-18}$ & m$^3$ & -\\
     Superfluid helium mass & $m_{\mathrm{He}}=V_{\mathrm{He}}\times \rho_{\mathrm{He}}$ &  $6 \times 10^{-16}$ & kg & -\\
     \hline
     \textbf{Device thermal properties}\\
       Superfluid latent heat of vaporization & $L_H$ & 17.5 & kJ/kg & \cite{donnelly_observed_1998} \\
   Estimated superfluid temperature & $T_{sf}$ & 500 & mK & FEM\\
      He II entropy @500 mK  & S & 0.8  & J/kg/K & \cite{donnelly_observed_1998}\\
      Cryostat temperature & $T_0$ &   100 & mK & meas.\\
      Silicon density & $\rho_{Si}$ & 2330 & kg/m$^3$ & \\
       Superfluid $^4$He density   & $\rho_{\mathrm{He}}$ & 145 & kg/m$^3$ & \cite{donnelly_observed_1998}\\
       Surface tension of superfluid $^4$He & $\sigma$ &  $3.54\times10^{-4}$ &  N/m & \cite{donnelly_observed_1998}\\
         \hline
    \end{tabular}
    \caption{Physical parameters. FEM: Finite Element modelling. SEM: Scanning electron microscope. Meas.: measurement. }
    \label{Table_physical_params}
\end{table}

\subsubsection{Influence of surface curvature}
\label{subsectioninfluenceofsurfacecurvature}

Surface tension is typically treated as if on a planar geometry: this is reasonable provided that the curvature of the geometry is insignificant compared to the wavelength. In the case of shallow waves on a slowly curving geometry, the main effect of surface tension is an effective modification to the restoring force.

In the case of the wave flume described here, the steps for showing this are walked through below.

To check this, one can work out the surface tension for a cylindrically symmetric surface $r(z)$, to see how it differs from a plane wave.  In the planar geometry, the curvature $\kappa_x$ along the wave propagation direction and the transverse curvature $k_y$ define the surface tension, as seen in the Young Laplace equation.

\begin{equation}
    \Delta p = \sigma \left( \frac{1}{R_x}+\frac{1}{R_y}\right)\label{eqn:YoungLaplace}
\end{equation}


It is straightforward to express the Young Laplace more generally via the mean curvature, which is a common differential geometry metric \cite{gray2017modern}). 

\begin{equation}
    \Delta p = 2 H \sigma
    \label{eqn:SurfaceTensionCurvature}
\end{equation}

Defining a cylindrical coordinate system with $\hat{z}$ bring the propagation direction, $\hat{\theta}$ being the axial direction, and $\hat{r}$ being the radial direction. This gives a total curvature \ref{eqn:SurfaceTensionCylindricalSurfac}.

\begin{equation}
    2 H = \dfrac{r r_{zz} - (r_z^2+1)}{r \left(r_z^2 +1 \right)^{\frac{3}{2}}}
    \label{eqn:CylindricalCurvature1}
\end{equation}

\begin{equation}
    2 H = \dfrac{r_{zz}}{\left(r_z^2 +1 \right)^{\frac{3}{2}}} + \dfrac{ -1}{r \sqrt{r_z^2 +1}}
    \label{eqn:SurfaceTensionCylindricalSurfac}
\end{equation}

Note that the first term of this is precisely what is expected in a one-dimensional wavetank. For large wavelengths, the second term reduces to what is expected for a cylindrical surface of radius $r$. 

Taking a wave profile for a thin film of superfluid on a cylinder of radius $R$: $r(z) = R + \eta(z)$ allows us to examine surface tension in a fluid $\eta(z)$. To do this we apply \ref{eqn:SurfaceTensionCurvature} and \ref{eqn:SurfaceTensionCylindricalSurfac} to the typical Cartesian hydrodynamic Euler equations discussed in Appendix \ref{methodssectionhydrodynamicmodelling}, which yields 
\begin{equation}
\phi_t + g \eta + \dfrac{\sigma}{\rho} \left(\dfrac{\eta_{zz}}{\left(\eta_z^2 +1 \right)^{\frac{3}{2}}} + \frac{ -1}{(R+\eta) \sqrt{\eta_z^2 +1}} \right) = 0. 
\end{equation}
Note that an effective gravity term can be obtained $g_0 = \sqrt{\tfrac{3 \alpha}{h^4}} - \tfrac{\sigma}{\rho R^2}$ for comparisons with standard water gravity waves, where the gravitational restoring force has been replaced with the van der Waals restoring force.

In the shallow wave limit (and for a thin film), the following dispersion relationship is obtained (for small amplitude waves). 

\begin{equation}
    \omega^2 = \left( \left(\sqrt{\dfrac{3 \alpha}{h^4}} - \dfrac{\sigma}{\rho R^2} \right) k + \dfrac{\sigma}{\rho} k^3 \right) \tanh(k h)
    \label{eqn:Dispersion}
\end{equation}

The Cartesian hydrodynamic Euler equation was used here for simplicity. This is reasonable when the radius of the cylinder is large compared to the depth and amplitude of the wave. Otherwise, the hydrodynamic Euler equation will need to be solved in cylindrical coordinates, and will yield solutions of the form of modified Bessel functions rather than exponential functions.

\subsection{Influence of dispersion and nonlinearity}
\label{sectionappendixinfluenceofdispersionandnonlinearity}

\begin{figure}[t]
    \centering
    \includegraphics[width=\textwidth]{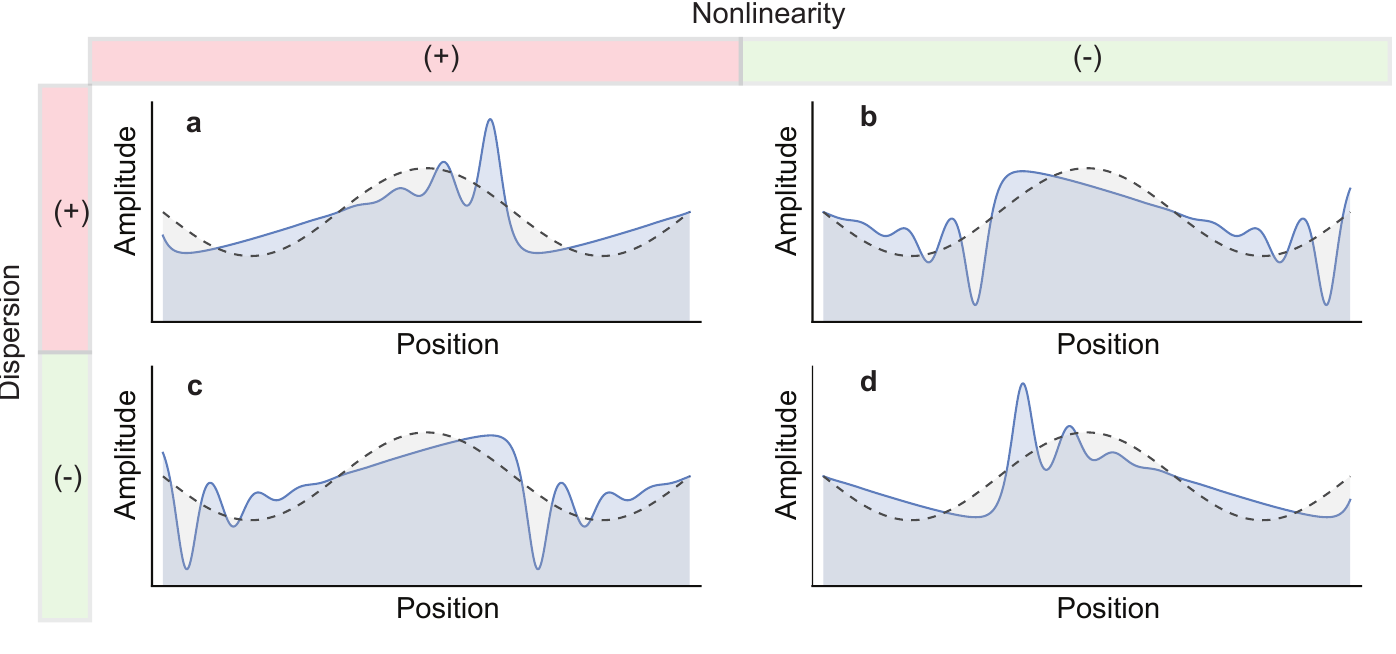}
    \caption{\textbf{Dispersion-nonlinearity matrix.} \textbf{a} Both nonlinear and dispersive terms are positive: ($\alpha =+6;\, \beta=+4$). This corresponds to the case of shallow water waves~\cite{trillo_experimental_2016}. \textbf{b} The sign of the nonlinearity is negative, while the dispersion is positive ($\alpha =-6;\, \beta=+4$).  This corresponds to the case explored in this work. \textbf{c} The nonlinearity is positive, while the dispersion is negative (anomalous dispersion): ($\alpha =+6;\, \beta=-4$). This is representative of the case of  capillary solitons in gravity-capillary waves \cite{perrard_capillary_2015}.  \textbf{d} Both nonlinearity and dispersion are negative ($\alpha =-6;\, \beta=-4$). This case corresponds to the case of  third sound in  thick superfluid films, where capillary effects dominate the dispersion, as predicted by Nakajima et al.~\cite{nakajima_solitary_1980}.  The dashed gray outline represents the initial sinusoidal wave profile $u(x,t=0)=40 \sin(x)$.  }
    \label{fig:nonlinearitydispersionmatrix}
\end{figure}

Solitons emerge due to the precise balancing of nonlinear and dispersive effects in the medium~\cite{onorato_rogue_2016,ellis_third_1992,zhang_optomechanical_2021}. The shape of the soliton solutions therefore depends on the sign (and magnitude) of both the dispersion and nonlinearity, with solutions switching from `dark' to `bright' (`hot' to `cold') solitons depending on the conditions which the wave experiences in the medium. 
This is illustrated in Fig.~\ref{fig:nonlinearitydispersionmatrix}, where we plot the evolving shape of an initially sinusoidal wave profile (dashed gray), whose evolution is governed by the KdV equation:
\begin{equation}
          u_t+ \alpha u\,u_x+ \beta u_{xxx}=0.
          \label{Eqkdvdispersionnonlinearitymatrix}
\end{equation}
Here the coefficients $\alpha$ and $\beta$ respectively quantify the nonlinearity and dispersion. Fig.~\ref{fig:nonlinearitydispersionmatrix}(a-d) plots the evolution of Eq.~\ref{Eqkdvdispersionnonlinearitymatrix} after the same time evolution ($t=0.006$) under conditions of varying nonlinearity and dispersion, with $\alpha$ and $\beta$ respectively taking the values ($\alpha =+6;\, \beta=+4$); ($\alpha =-6;\, \beta=+4$); ($\alpha =+6;\, \beta=-4$) and ($\alpha =-6;\, \beta=-4$). 

Figure~\ref{fig:nonlinearitydispersionmatrix}(a) shows the wave evolution in the case where both nonlinear and dispersive terms are positive: ($\alpha =+6;\, \beta=+4$). This corresponds to the case of shallow water waves~\cite{trillo_experimental_2016}, where the crests move faster than the troughs and the group velocity is reduced at higher wavenumbers (normal dispersion). This is the regime in which multisoliton fission was recently observed in a hundred meter-scale wave flume~\cite{trillo_experimental_2016}. Solitons manifest as a propagating elevation above the fluid baseline.

Figure~\ref{fig:nonlinearitydispersionmatrix}(b) is representative of the case observed in this work: ($\alpha =-6;\, \beta=+4$). Here the sign of the nonlinearity is reversed compared to the shallow water case (with troughs progressing faster than peaks) due to the nonlinearity of the van der Waals interaction, but the normal dispersion is maintained. Indeed, the normal dispersion introduced by the phononic crystal exceeds the anomalous dispersion due to capillary effects (see section~\ref{sectionsuppacousticbraggmirrordispersion}).  In this case, waves break backwards and solitons manifest as a propagating depression below the fluid baseline (so-called `hot solitons' in the superfluid context). This regime had been predicted to occur in very thin superfluid films (thickness $h_0\simeq 1$ nm) for which the dispersive influence of surface tension could be neglected~\cite{nakajima_solitons_1980}. A key difference with this work is the characteristic length scale of the solitons, which is on the order of micrometers here because of the larger dispersion, vs only on the order of the nanometer film thickness in Ref.~\cite{nakajima_solitons_1980}.

Figure~\ref{fig:nonlinearitydispersionmatrix}(c) shows the wave evolution in the case ($\alpha =+6;\, \beta=-4$), where the nonlinear term is positive (peaks propagate faster than troughs), while the  dispersive term is  negative. This corresponds to anomalous dispersion where the higher wave numbers propagate with a larger group velocity. This case is representative of  capillary solitons in gravity-capillary waves \cite{perrard_capillary_2015}.

Finally, Figure~\ref{fig:nonlinearitydispersionmatrix}(d) shows the wave evolution in the case ($\alpha =-6;\, \beta=-4$). This corresponds to peaks progressing slower than troughs (as is the case with a nonlinear van der Waals restoring force) and anomalous dispersion. This case has been predicted to occur in saturated (thick) superfluid helium films by Nakajima et al.~\cite{nakajima_solitary_1980}, where capillary effects dominate the dispersion. The resulting solitons manifest, as in the shallow water gravity case (a), as traveling elevations above the baseline---so-called `cold' solitons in the superfluid context. 

 \end{document}